\documentclass[review, 1p, number,sort&compress]{elsarticle}

\usepackage{amssymb}
\usepackage{lineno}

\journal{Nuclear Instruments and Methods A}

\begin{document}

\vspace*{-2cm}

\begin{frontmatter}


 \title{Methods for optical calibration of the BigBite hadron spectrometer}

\author[address_22]{M.~Mihovilovi\v{c}}
\author[address_22,address_27]{S.~\v{S}irca}
\author[address_1]{K.~Allada}
\author[address_2]{B.~D.~Anderson}
\author[address_3]{J.~R.~M.~Annand}
\author[address_4]{T.~Averett}
\author[address_1]{A.~Camsonne}
\author[address_9]{R.~W.~Chan}
\author[address_1]{J.-P.~Chen}
\author[address_9]{K.~Chirapatpimol}
\author[address_1]{C.~W.~de~Jager}
\author[address_15]{S.~Gilad}
\author[address_3]{D.~J.~Hamilton}
\author[address_1]{J.-O.~Hansen}
\author[address_1]{D.~W.~Higinbotham\corref{cor1}}
\cortext[cor1]{Corresponding author. Tel.: +1-757-269-7851; fax: +1-757-269-7881.}
\ead{doug@jlab.org}
\author[address_15]{J.~Huang}
\author[address_17]{X.~Jiang}
\author[address_9]{G.~Jin}
\author[address_10]{W.~Korsch}
\author[address_1]{J.~J.~LeRose}
\author[address_9]{R.~A.~Lindgren}
\author[address_9]{N.~Liyanage}
\author[address_2]{E.~Long}
\author[address_1]{R.~Michaels}
\author[address_1]{B.~Moffit}
\author[address_7]{P.~Monaghan}
\author[address_9]{V.~Nelyubin}
\author[address_9]{B.~E.~Norum}
\author[address_24]{E.~Piasetzky}
\author[address_14]{X.~Qian}
\author[address_14]{Y.~Qiang}
\author[address_9]{S.~Riordan}
\author[address_30]{G.~Ron}
\author[address_31]{G.~Rosner}
\author[address_1]{B.~Sawatzky}
\author[address_9]{M.~Shabestari}
\author[address_26]{A.~Shahinyan}
\author[address_24]{R.~Shneor}
\author[address_29]{R.~Subedi}
\author[address_1,address_15]{V.~Sulkosky}
\author[address_1]{J.~W.~Watson}
%%%\author[address_1]{B.~Wojtsekhowski}
\author[address_11]{Y.-W.~Zhang}

 %%\fntext[label3]{Jozef ....}

\address[address_22]{Jo\v{z}ef Stefan Institute, 1000 Ljubljana, Slovenia}
\address[address_27]{University of Ljubljana, 1000 Ljubljana, Slovenia}
\address[address_1]{Thomas Jefferson National Accelerator Facility, Newport News, VA 23606, USA}
\address[address_2]{Kent State University, Kent, OH, 44242, USA} 
\address[address_3]{Glasgow University, Glasgow, G12 8QQ, Scotland, United Kingdom} 
\address[address_4]{College of William and Mary, Williamsburg, VA, 23187, USA}
\address[address_9]{University of Virginia, Charlottesville, VA, 22908, USA}
\address[address_15]{Massachusetts Institute of Technology, Cambridge, MA, 02139, USA}
\address[address_17]{Los Alamos National Laboratory, Los Alamos, NM, 87545, USA}
\address[address_10]{University of Kentucky, Lexington, KY, 40506, USA}
\address[address_7]{Hampton University , Hampton, VA, 23668, USA}
\address[address_24]{Tel Aviv University, Tel Aviv 69978, Israel}
\address[address_14]{Duke University, Durham, NC, 27708, USA}
\address[address_30]{Hebrew University of Jerusalem, Jerusalem, Israel}
\address[address_31]{GSI, 64291 Darmstadt, Germany}
\address[address_26]{Yerevan Physics Institute, Yerevan, Armenia}
\address[address_29]{George Washington University, Washington, D.C., 20052, USA}
\address[address_11]{Rutgers University, New Brunswick, NJ, 08901, USA}

%\address[address_30]{Norfolk State University, Norfolk, VA, 23504, USA}
%\address[address_5]{Florida International University, Miami, FL, 33181, USA}
%\address[address_6]{Old Dominion University, Norfolk, VA, 23508, USA}
%\address[address_8]{Universite di Bari, Bari, 70121 Italy} 
%\address[address_12]{Temple University, Philadelphia, PA, 19122, USA}
%\address[address_13]{Istituto Nazionale Di Fisica Nucleare, INFN/Sanita, Roma, Italy} 
%\address[address_16]{Christopher Newport University, Newport News, VA, 23606, USA}
%\address[address_18]{Seoul National University, Seoul, Korea}
%\address[address_19]{Ohio University, Athens, OH, 45701, USA}
%\address[address_20]{Huangshan University, China}
%\address[address_21]{Lanzhou University, Lanzhou, Gansu, 730000, China}
%\address[address_23]{Carnegie Mellon University, Pittsburgh, PA, 15213, USA}
%\address[address_25]{Northern Michigan University, Marquette, MI, 49855, USA}
%\address[address_28]{Longwood University, Farmville, VA, 23909, USA}

\vspace*{-19mm}

\begin{abstract} 
The techniques for optical calibration of Jefferson Lab's 
large-acceptance magnetic hadron spectrometer, BigBite,  
have been examined.
The most consistent and stable results were obtained
by using a method based on singular value decomposition.
In spite of the complexity of the optics, the particles'
positions and momenta at the target have been precisely
reconstructed from the coordinates measured in the detectors
by means of a single back-tracing matrix.  The technique
is applicable to any similar magnetic spectrometer
and any particle type.  For $0.55\,\mathrm{GeV}/c$ protons,
we have established a vertex resolution of $1.2\,\mathrm{cm}$,
angular resolutions of $7\,\mathrm{mrad}$ and $13\,\mathrm{mrad}$
(in-plane and out-of-plane, respectively), and a relative
momentum resolution of $1.6\,\mathrm{\%}$.
\end{abstract}


\end{frontmatter}

%%\linenumbers


\section{Introduction}

One of the recent acquisitions in experimental Hall A
of the Thomas Jefferson National Accelerator Facility (TJNAF) 
is the BigBite spectrometer.  It was previously
used at the NIKHEF facility for the detection of electrons
\cite{lange-general,lange-optics}.  At Jefferson Lab,
BigBite has been re-implemented as a versatile spectrometer
that can be instrumented with various detector packages
optimized for the particular requirements of the experiments.
BigBite complements the High-Resolution Spectrometers,
which are part of the standard equipment of Hall~A \cite{alcorn}.
Adding BigBite allows one to devise more flexible experimental
setups involving double- and even triple-coincidence measurements.

In 2005, the BigBite spectrometer was first used in Hall~A 
as the hadron arm in the E01-015 experiment, which investigated
nucleon-nucleon short-range correlations \cite{subedi, shneor}.
In 2006, it was instrumented as the electron arm for the measurement
of the neutron electric form factor (experiment E02-013 \cite{riordan}). 
In 2008 and 2009, it has been used in two large groups of experiments
spanning a broad range of physics topics.  We studied near-threshold
neutral pion production on protons (experiment E04-007 \cite{e04007})
and measured single-spin asymmetries in semi-inclusive pion electro-production 
on polarized $^3\mathrm{He}$ (experiments E06-010 and E06-011
\cite{qian11,huang11}).  In the same period,
we also measured parallel and perpendicular asymmetries
on polarized $^3\mathrm{He}$ in order to extract
the $g_2^\mathrm{n}$ polarized structure function in
the deep-inelastic regime (experiment E06-014 \cite{e06014}),
and measured double-polarization asymmetries in the quasi-elastic processes 
${}^3\vec{\mathrm{He}}(\vec \mathrm{e},\mathrm{e}'\mathrm{d})$,
${}^3\vec{\mathrm{He}}(\vec \mathrm{e},\mathrm{e}'\mathrm{p})$, and
${}^3\vec{\mathrm{He}}(\vec \mathrm{e},\mathrm{e}'\mathrm{n})$ 
(experiments E05-102 and E08-005 \cite{e05102, e08005}). 
In 2011, the investigation of short-range 
correlations has been continued in the E07-006 experiment \cite{e07006}
exploring the repulsive part of the nucleon-nucleon interaction.

BigBite is a non-focusing spectrometer consisting of
a single dipole with large momentum and angular acceptances
(the details are presented in Section~\ref{sec:BB}).
The magnetic optics of such spectrometers tend to become
complicated towards the edges of their acceptances,
especially for the momentum and the dispersive angle.
It was not clear from the outset that particle momentum
and interaction vertex reconstruction could be accomplished
by using a single procedure for all momenta.

The calibration presented in this paper allows for
a full description of BigBite optics by means of a single
reconstruction matrix.  The method was developed and successfully
used with the data obtained in the E05-102 experiment with
the detector package configured for hadrons
(Section~\ref{sec:exp}), but it is applicable
to any magnetic spectrometer with a similar optical configuration
and any particle type. Various calibration procedures are 
discussed in Section~\ref{sec:cal}.

\section{The BigBite spectrometer}
\label{sec:BB}

\begin{figure}[!ht]
\begin{center}
\includegraphics[width=13cm]{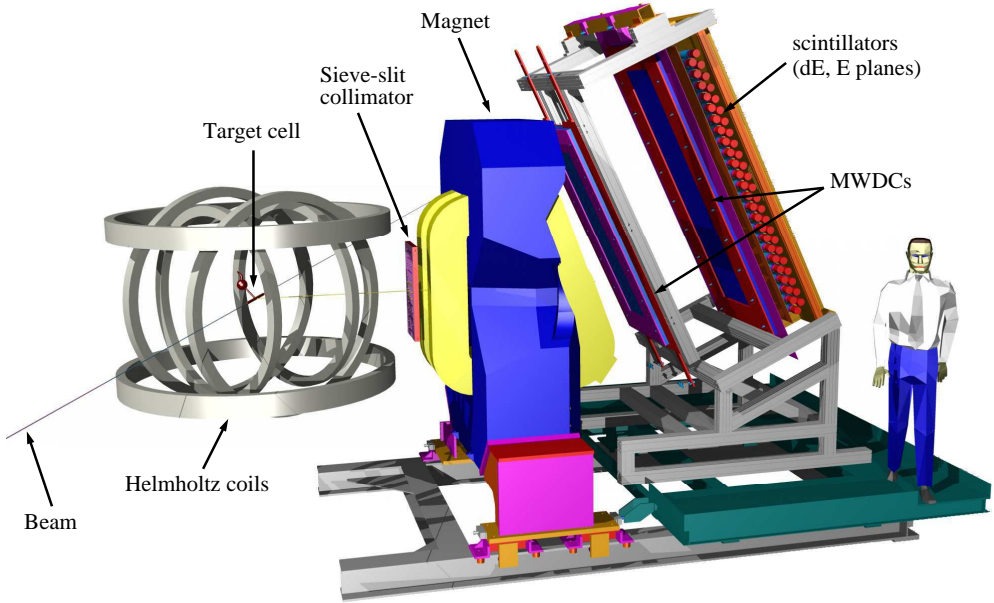}
\vspace*{-3mm}
\caption{The BigBite spectrometer on its support frame. 
BigBite consists of a dipole magnet, followed by the
detector package assembled from a pair of multi-wire drift 
chambers (MWDC) and two scintillator planes (dE and E). 
The directions of the incoming electron beam and the scattered
particles, the target cell, and the Helmholtz coil (holding field)
assembly are also shown.\label{BBSpectrometer}}
\end{center}
\end{figure}

The BigBite spectrometer \cite{lange-general} consists of a single
room-temperature dipole magnet, shown in Fig.~\ref{BBSpectrometer}.
Energizing the magnet with a current of $518\,\mathrm{A}$
results in a mean field density of $0.92\,\mathrm{T}$,
corresponding to a central momentum of $p_\mathrm{c}=0.5\,\mathrm{GeV}/c$
and a bending angle of $25^\circ$.  The magnet is coupled to
 a hadron detector package consisting of two
multi-wire drift chambers (MWDC) \cite{nilanga,ChanMSc}
for particle tracking and two planes of scintillation
detectors (denoted by dE and E) \cite{ShneorMSc} for triggering,
particle identification, and energy determination. 

Each MWDC consists of six planes of wires.  The wires
in the first two planes are oriented at an angle of $60^\circ$
with respect to the dispersive direction, while the wires
in the third and fourth plane are aligned horizontally.
The wires of the last two planes are oriented at $-60^\circ$.
Each wire plane in the first and the second MWDC
contains $141$ and $200$ wires, respectively.  
The spacing between the wires in all planes is  $1\,\mathrm{cm}$.
The intrinsic spatial resolution of the MWDCs is about
$100\,\mu\mathrm{m}$ and $200\,\mu\mathrm{m}$ for the dispersive
and non-dispersive coordinates, respectively, and about 
$0.15\,\mathrm{mrad}$ and $0.35\,\mathrm{mrad}$ for the dispersive
and non-dispersive angles, respectively.

The dE- and E-planes (also called the trigger planes)
each consist of $24$ plastic scintillator bars.
The bars are $50\,\mathrm{cm}$ long and $8.6\,\mathrm{cm}$ wide.
For the dE-plane, thinner bars ($0.3\,\mathrm{cm}$) were used
to detect low-energy particles, while for the E-plane,
a thickness of $3\,\mathrm{cm}$ was chosen to allow for
the detection of more energetic particles.  The light pulses
in each bar were detected by photomultiplier tubes mounted
at each end of the bar.  To double the spatial and momentum resolution,
the bars in the E-plane are offset from those in the dE-plane
by one half of the bar width ($4.3\,\mathrm{cm}$).

\section{Experimental details and data}
\label{sec:exp}

The E05-102 experiment was performed in Hall~A~\cite{alcorn}
at Jefferson Lab.  In the experiment, a polarized ${}^3\mathrm{He}$
target was used in conjunction with the polarized continuous-wave
electron beam.  Scattered electrons were detected
by the left High Resolution Spectrometer (HRS) in coincidence
with protons and deuterons that were detected by BigBite.
A variety of kinematic settings were employed
(Table~\ref{table_kinematics}), with the momentum-transfer 
vector $\vec{q}$ pointing towards BigBite.  This ensured
that the protons and deuterons from elastic and quasi-elastic
scattering were always within its acceptance.

\begin{table}[!ht]
\caption{Kinematics settings of the E05-102 experiment
(the incoming electron energy $E_{\mathrm{beam}}$ and
the angles of the HRS and BigBite spectrometers with respect
to the beam direction).\label{table_kinematics}}
\begin{center}
\begin{tabular}{lccc}
\hline

\hline 

\hline
 Setting  & $E_{\mathrm{beam}}$ 
          & \multicolumn{2}{c}{Scattering angle }  \\[-6pt] 
 label    & [$\mathrm{GeV}/c$] & HRS [${}^\circ$] & BB [${}^\circ$]\\ 

\hline 
$1$-pass    & $1.245$ & $17.0$ & $-74.0$ \\
\hline
$2$-pass    & $2.425$ & $12.5$ & $-75.0$ \\[-6pt]
            &         & $14.5$ & $-82.0$ \\
\hline
$3$-pass    & $3.606$ & $12.5$ & $-75.0$ \\[-6pt]
            &         & $17.0$ & $-74.0$ \\
\hline

\hline

\hline
\end{tabular}
\end{center}
\end{table}

\begin{figure}[!ht]
\begin{center}
\includegraphics[width=0.85\textwidth]{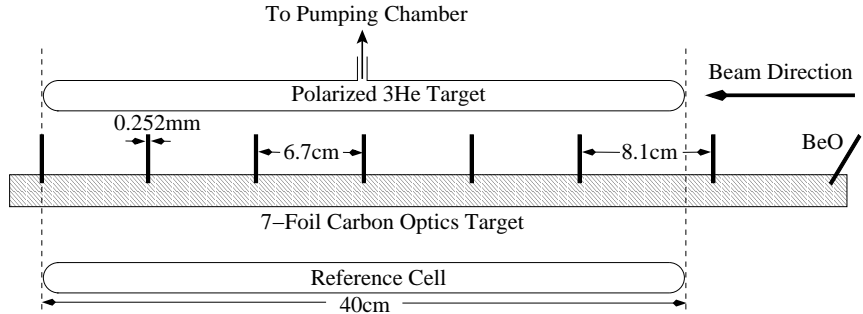}
\caption{The target system including the polarized
${}^3\mathrm{He}$ cell at the top, the multi-foil carbon optics target,
and the reference cell at the bottom.  The slanted BeO foil
is used for visual inspection of the beam impact point.\label{TGSYSTEM}}
\end{center}
\end{figure}

The core component of the polarized ${}^3\mathrm{He}$ target
was a pressurized cylindrical glass cell with a length
of $40\,\mathrm{cm}$ and a diameter of $1.9\,\mathrm{cm}$
(see Fig.~\ref{TGSYSTEM}).  The thickness
of the glass cylinder was $1.7\,\mathrm{mm}$, while the thickness
of the end windows was $140\,\mu\mathrm{m}$.  The gas in the cell
was polarized to approximately $60\,\%$ by hybrid spin-exchange
optical pumping \cite{he3a,he3b} driven by an infra-red laser system.
The direction of the nuclear polarization was maintained 
by three pairs of Helmholtz coils surrounding the cell.

\goodbreak

In addition to the ${}^3\mathrm{He}$ helium target,
a $40\,\mathrm{cm}$-long multi-foil carbon target
was used for calibration, as described below.
It consists of seven $0.252\,\mathrm{mm}$-thick carbon foils
mounted to a plastic frame (Fig.~\ref{TGSYSTEM})
which are preceded by a single slanted BeO foil for beam positioning.
Below the multi-foil target, a dummy (reference) cell was installed
that could be either evacuated or filled with hydrogen, deuterium, 
unpolarized helium-3 or nitrogen. 

\begin{figure}[!ht]
\begin{center}
\includegraphics[height=8cm]{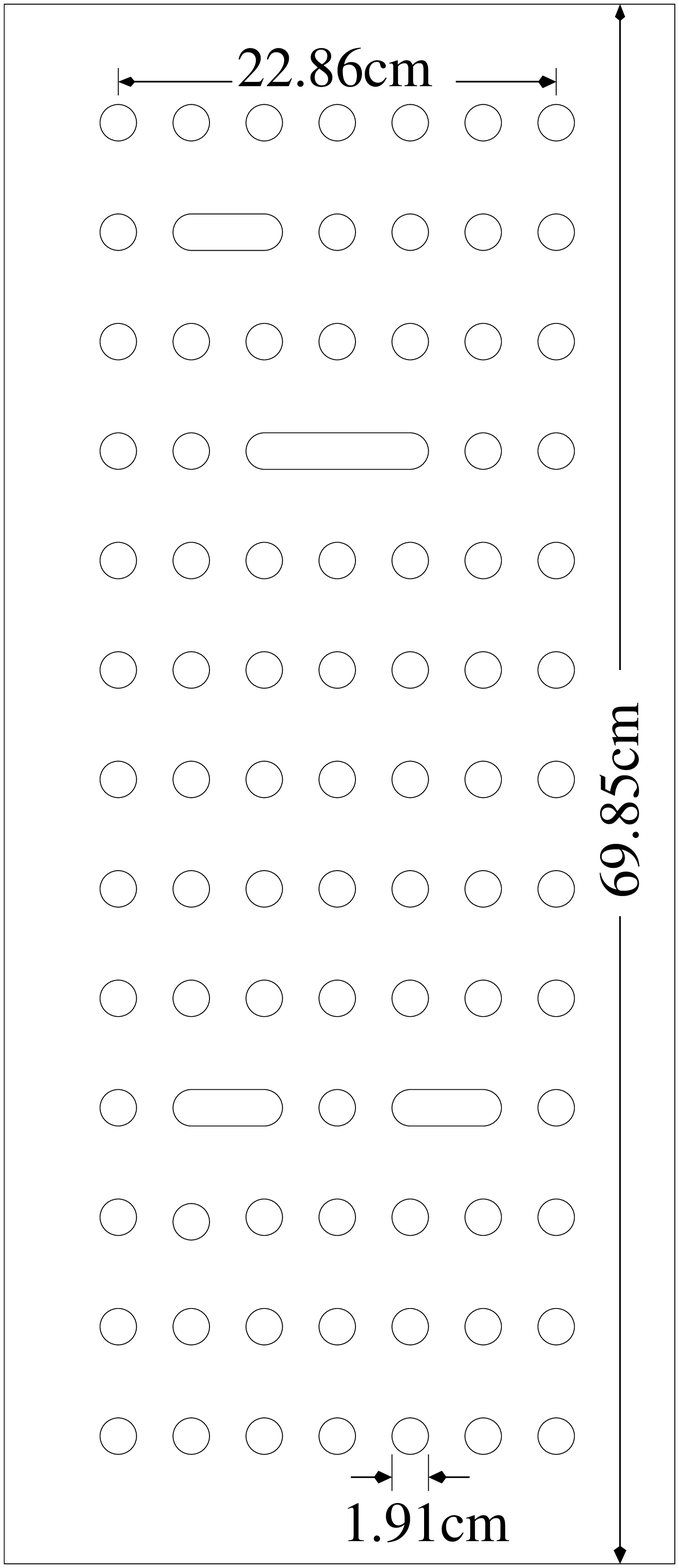}
\includegraphics[height=8cm]{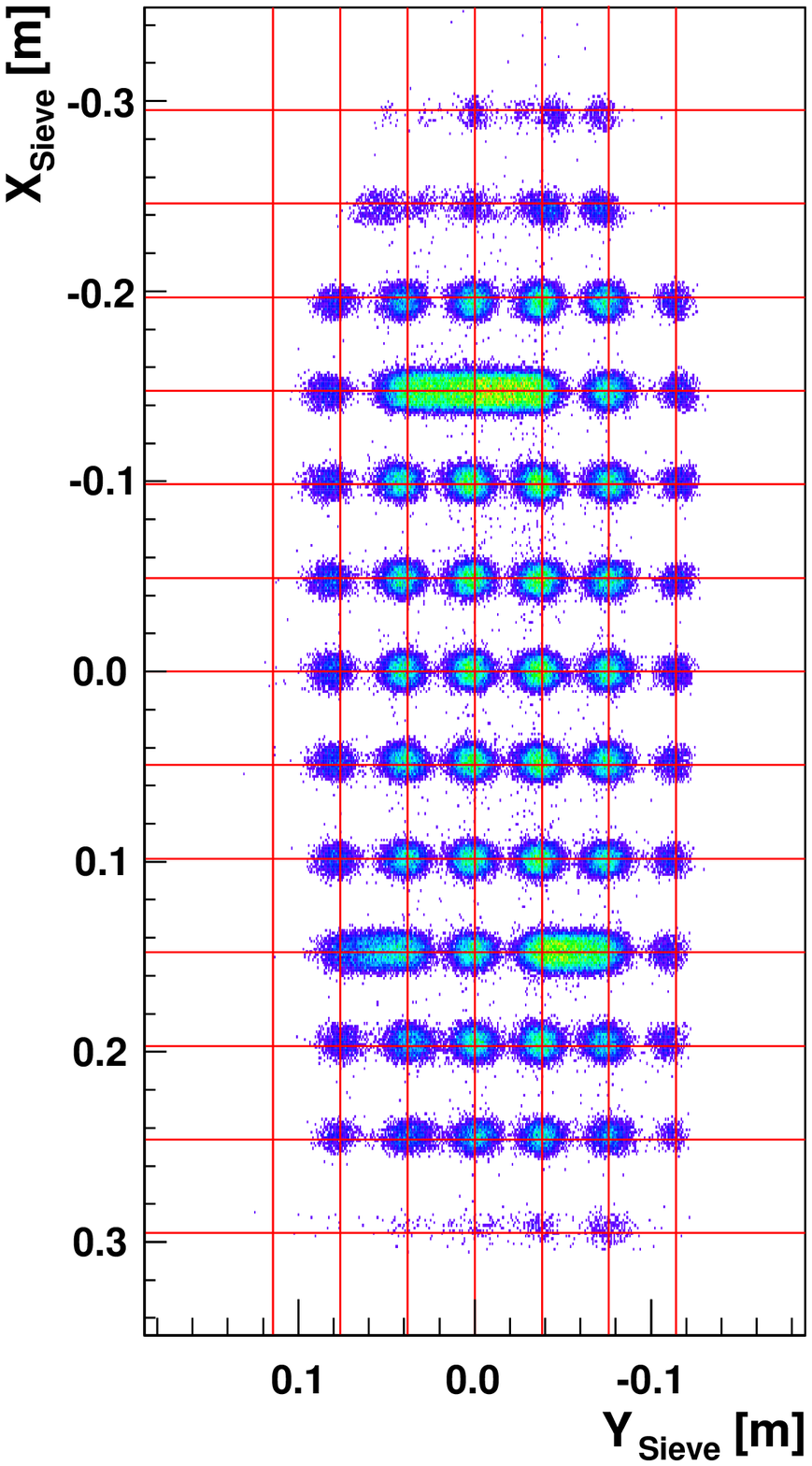}
\includegraphics[height=8cm]{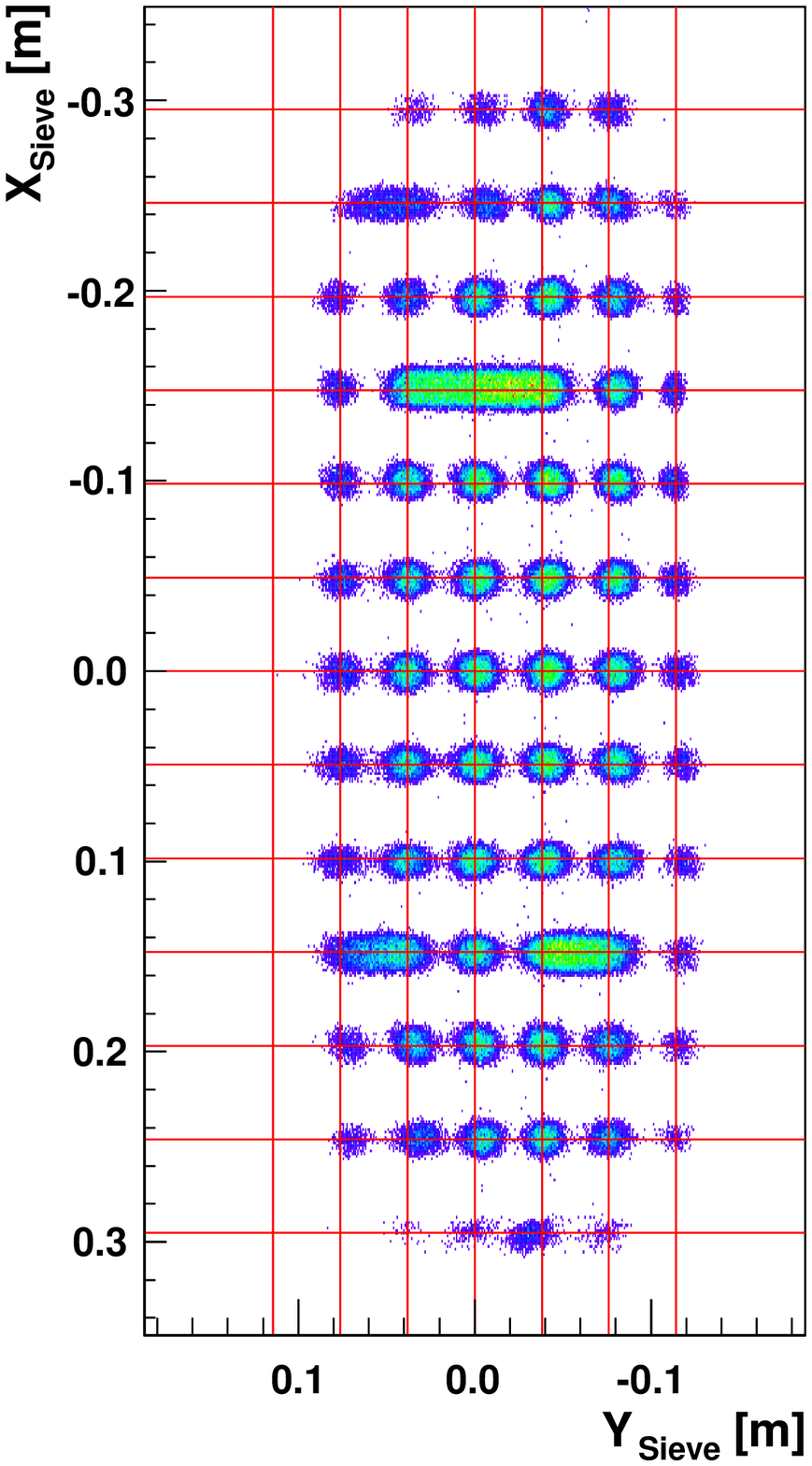}
\caption{[Left] Schematics of the BigBite sieve-slit collimator.
[Center, Right] Sieve pattern reconstruction by using the simplex method
and the SVD, respectively (see subsubsection~\ref{sub:nmsvd}).
The SVD technique resolves more holes and yields a much clearer pattern.
The holes at the left edge are missing due to geometrical obstacles
between the target and BigBite.\label{BBSieve}}
\end{center}
\end{figure}

For the optics calibration of BigBite, a special set
of measurements was performed with a $4\,\mathrm{cm}$-thick
lead sieve-slit collimator positioned at the entrance
to the BigBite magnet (see Fig.~\ref{BBSpectrometer}).
The sieve-slit collimator has $82$ circular holes that are
almost uniformly positioned over the whole acceptance
of the spectrometer, Fig.~\ref{BBSieve} (left).
The collimator also contains four elongated holes used
to remove ambiguities in horizontal and vertical orientations
and to allow for easier identification of the hole projections
at the detector package. 

Prior to any optics analysis, a series of cuts were applied 
to the collected calibration data to eliminate backgrounds.
A HRS-BigBite coincidence trigger system was used to acquire
electron-proton and electron-deuteron coincidences,
at typical rates between $700\,\mathrm{Hz}$ and $1\,\mathrm{kHz}$.
True coincidences were selected by applying a cut on the raw
coincidence time.  The backgrounds were further reduced
by PID and HRS acceptance cuts.  Finally, only those events
that produce consistent hits in all BigBite detectors,
and could consequently be joined to form single particle tracks,
were selected.

\section{Methods of optical calibration}
\label{sec:cal}

The purpose of optical calibration is to establish the mapping
between the detector variables that are measured directly,
and the target variables corresponding to the actual physical 
quantities describing the particle at the reaction vertex.
In the MWDCs, two position coordinates ($x_{\mathrm{Det}}$
and $y_{\mathrm{Det}}$) and two angles ($\theta_{\mathrm{Det}}$
and $\phi_{\mathrm{Det}}$) are measured.  From this information,
we wish to reconstruct the location of the interaction vertex 
($y_{\mathrm{Tg}}$), the in-plane and out-of-plane scattering
angles ($\phi_{\mathrm{Tg}}$ and $\theta_{\mathrm{Tg}}$),
and the particle momentum relative to the central momentum
($\delta_{\mathrm{Tg}}=(p_\mathrm{Tg}-p_\mathrm{c})/p_\mathrm{c}$).
This can be done in many ways.  We have considered an analytical model
as well as a more sophisticated approach based on transport-matrix
formalism, with several means to estimate the reliability
of the results and the stability of the algorithms.

Quasi-elastic protons from scattering on the multi-foil carbon
target were used to calibrate $y_\mathrm{Tg}$; the same target
was also used to calibrate $\theta_\mathrm{Tg}$ and $\phi_\mathrm{Tg}$
when the sieve-slit collimator was in place.  In turn, elastic protons
and deuterons (from hydrogen and deuterium targets) were used
to calibrate $\theta_\mathrm{Tg}$, $\phi_\mathrm{Tg}$, and
$\delta_{\mathrm{Tg}}$.
The $\delta_{\mathrm{Tg}}$ matrix elements could also be determined
by quasi-elastic events from $^3\mathrm{He}$ under the assumption
that the energy losses are well understood.

\begin{figure}[hbtp]
\begin{center}
\includegraphics[width=15cm]{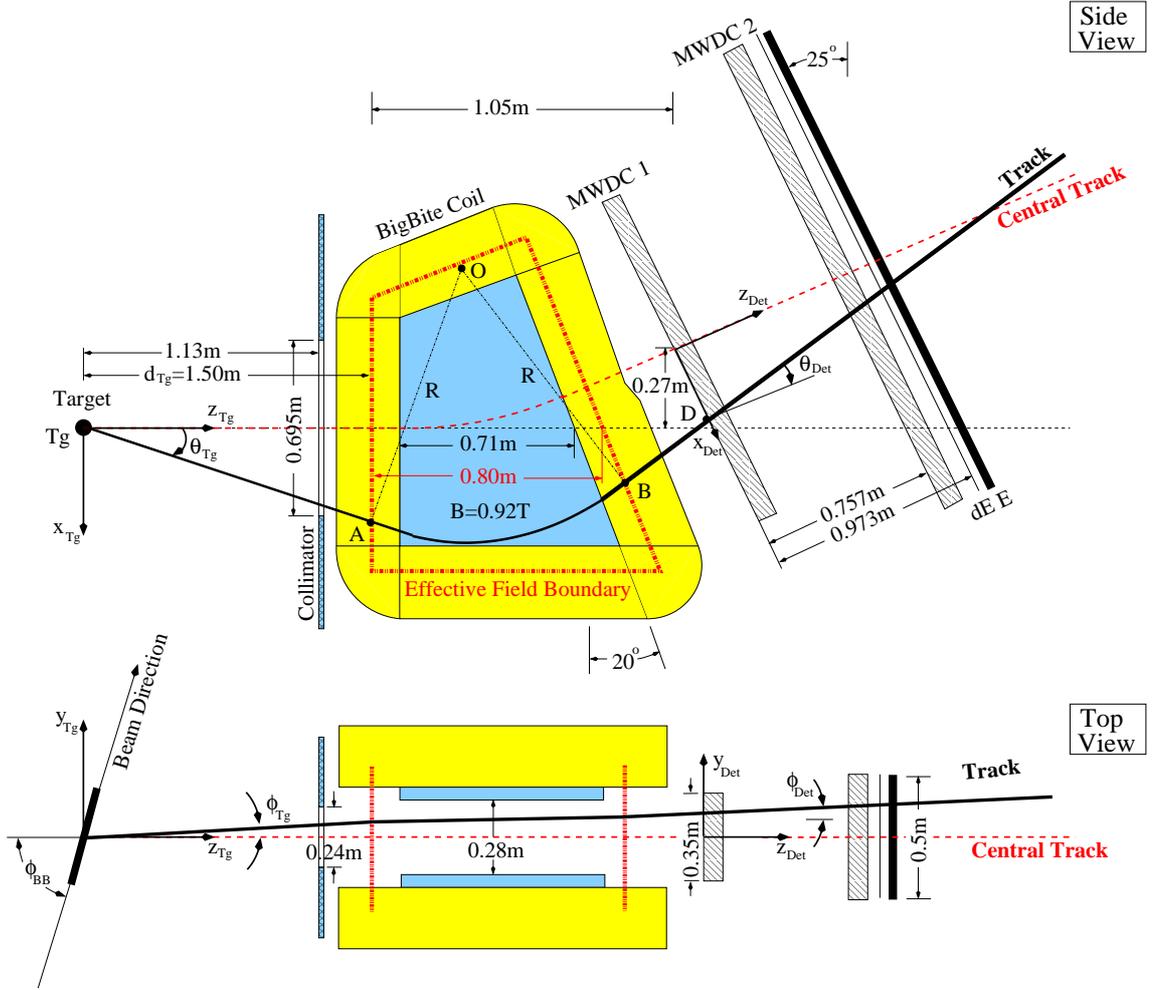} 
\caption{The schematic of the dispersive (top) and non-dispersive
(bottom) planes of the BigBite spectrometer.  Small angular
deflections in the non-dispersive plane occur if the particle
trajectory is not perpendicular to the effective field boundary
\cite{lange-optics,penner,brown}.  At the entrance to the magnet,
they are at most $18\,\mathrm{mrad}$ (close to the acceptance
boundaries in the dispersive direction).  At the exit field
boundary, the effect acts in the opposite sense and partially
cancels the deflection at the entrance.\label{BBScheme}}
\end{center}
\end{figure}

\subsection{The analytical model}

The magnetic field of the BigBite magnet is oriented in the
$y_{\mathrm{Tg}}$ direction (see Fig.~\ref{BBScheme}).
Field mapping has shown \cite{lange-general} that the field density
is almost constant inside the magnet, with fringe fields
that decrease exponentially outside of the magnet.
In the analytical model, the true field was approximated
by a constant field within the effective field boundaries,
while edge effects were neglected.  Under these assumptions
all target coordinates were calculated by applying
a circular-arc approximation~\cite{ShneorPhD} of the track
inside the field: the particle transport was divided into
free motion (drift) in the $(y,z)$ plane and circular motion
in the $(x,z)$ plane (see Fig.~\ref{BBScheme}), described
by the Lorentz equation
$$
p_{y} = \mathrm{const} \>, \qquad p_{xz}= e R B_y \>. 
$$
To determine the momentum, the radius $R$ of the trajectory needs
to be calculated first.  This can be done by using the track
information obtained from the detector package, combined
with the geometrical properties of BigBite. 
A few reference points are needed, as shown in Fig.~\ref{BBScheme};
the point $\mathrm{Tg}$ represents the position of the particle
at the target, and $D$ corresponds to the point where the particle
hits the detector package.  The point $B$ at which the particle
exits the magnet is the intersection between the extrapolated
particle track through the detector package and the effective 
exit face of the magnet.  Similarly, the point $A$ lies
at the intersection of the effective entrance face
of the magnet and the particle track from the target.
The point $O$ is the center of the circular trajectory. 
In order for all these points to correspond to a single
particle  track through the spectrometer, the conditions
\begin{eqnarray}
   \overline{A\mathrm{{Tg}}} \perp \overline{AO} \>, \qquad 
   \overline{OB} \perp \overline{BD} \>, \qquad
   |\overline{AO}| = |\overline{BO}| = R \>, \nonumber
\end{eqnarray}
must be satisfied.  In the target coordinate system, this becomes
\begin{eqnarray}
   x_O &=& -\frac{d_{\mathrm{Tg}}}{(x_A-x_{\mathrm{Tg}})}(z_O-z_A) + x_A  
        = -\frac{z_D - z_B}{x_D - x_B}(z_O - z_B) 
        + x_B\>, \label{analitic.eq1}\\
   R^2 &=& (z_O - z_A)^2\left[1 + \left(\frac{d_{\mathrm{Tg}}}{x_A-x_{\mathrm{Tg}}} \right)^2 \right]  
        = (z_B - z_O)^2\left[1 
        + \left(\frac{z_D - z_B}{x_D - x_B} \right)^2 \right]\,.\label{analitic.eq2}
\end{eqnarray}
The coordinates $x_B$ and $z_B$ of $B$, and the coordinates $x_D$
and $z_D$ of $D$ can be directly calculated from the information 
obtained by the detector package.  The position of the target 
$(x_{\mathrm{Tg}},z_{\mathrm{Tg}})$ is known. Since only thin targets
are employed, $x_{\mathrm{Tg}}$ is set to zero. The coordinate
$z_A$ of $A$ corresponds to the known distance $d_{\mathrm{Tg}}$
between the target 
center and the effective field boundary at the entrance to the magnet. 
By expressing $z_O$ from Eq.~(\ref{analitic.eq1}) and inserting
it into Eq.~(\ref{analitic.eq2}), an equation for $x_A$ is obtained
which has three complex solutions in general.  The physically
meaningful result for $x_A$ should be real and lie within
the effective field boundaries.  Two additional physical
constraints are applied.  The particle track should always
represent the shortest possible arc of the circle
(the arc between $A$ and $B$ in Fig.~\ref{BBScheme}).
Moreover, the track should bend according to the polarity
of the particle and orientation of the magnetic field.
From $x_A$, the radius $R$ and the momentum $p_{xz}$ 
can be calculated.  The particle flight path $l_{xz}$
in the $(x,z)$ plane can also be calculated by using
the cosine formula for the angle $\beta = \measuredangle AOB$,
\begin{eqnarray}
  l_{xz} &=&  \sqrt{x_A^2 + d_{\mathrm{Tg}}^2} + R\beta + \sqrt{(x_D-x_B)^2 
          + (z_D - z_B)^2}\>, \nonumber \\
  \cos \beta &=& \frac{(x_A-x_O)(x_B-x_O) 
          + (z_A-z_O)(z_B-z_O)}{R^2} \> . \nonumber
\end{eqnarray}
By using this information, all target coordinates can be expressed as
\begin{eqnarray}
  \phi_{\mathrm{Tg}} &=& \phi_{\mathrm{Det}} \>,\nonumber \\
  \theta_{\mathrm{Tg}} &=& \arctan\left(\frac{x_A}{d_{\mathrm{Tg}}}\right)  \>,\nonumber \\  
  y_{\mathrm{Tg}} &=& y_{\mathrm{Det}} 
    - l_{xz}\tan\phi_{\mathrm{Det}}\>,  \nonumber \\
  \delta_{\mathrm{Tg}} &=& \frac{p_{xz}}{p_\mathrm{c}}\frac{\sqrt{1
    +\tan^2\phi_{\mathrm{Tg}} + \tan^2\theta_{\mathrm{Tg}}}}{\sqrt{1
    + \tan^2\theta_{\mathrm{Tg}}}} - 1\>,\nonumber \\	
  L &=&l_{xz}\sqrt{1 + \tan^2\phi_{\mathrm{Tg}}}\>, \nonumber	
\end{eqnarray}
where $p_\mathrm{c}$ is the central momentum and $L$ is the 
total flight-path of the particle.   

\begin{figure}[!ht]
\begin{center}
\includegraphics[width=0.49\textwidth]{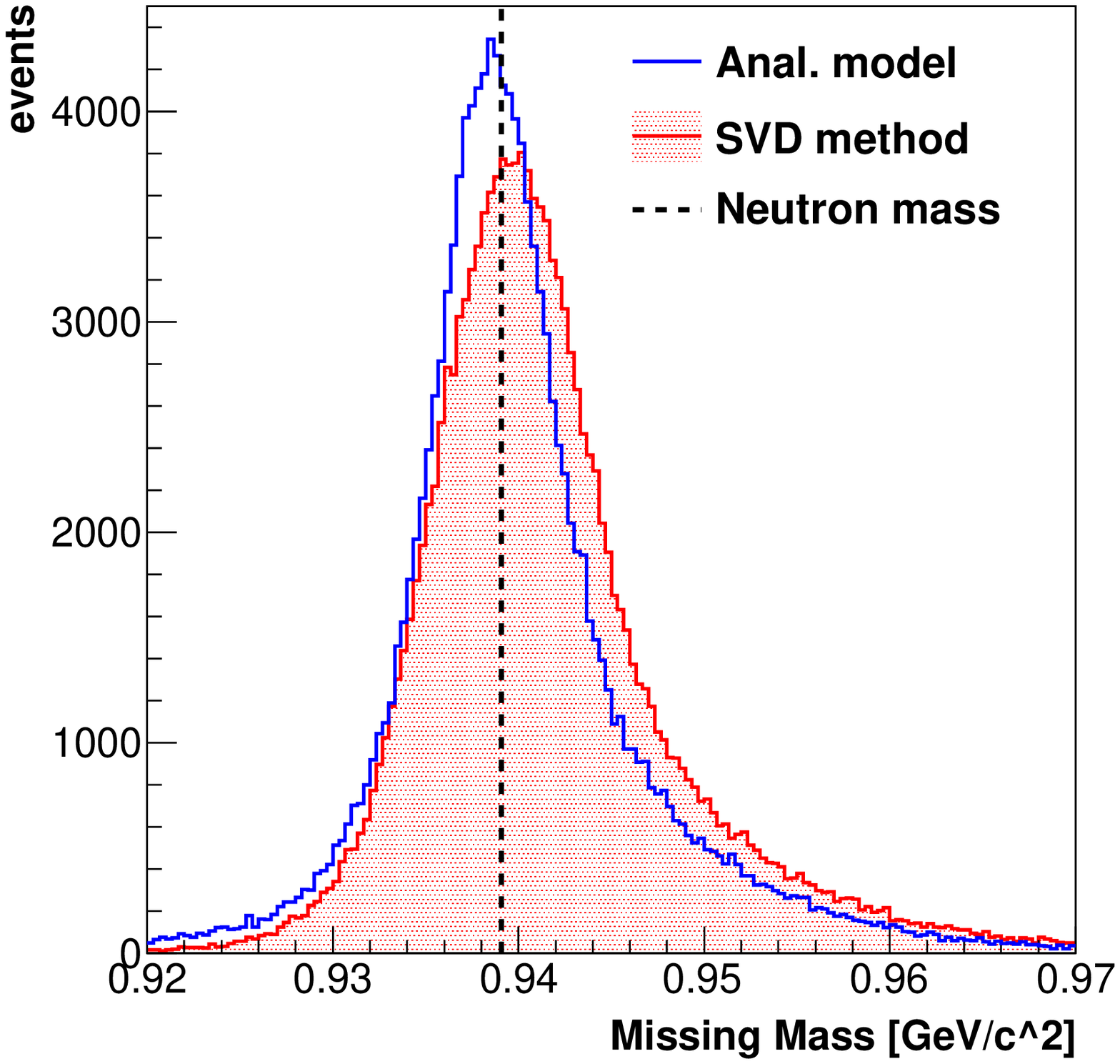}
\includegraphics[width=0.49\textwidth]{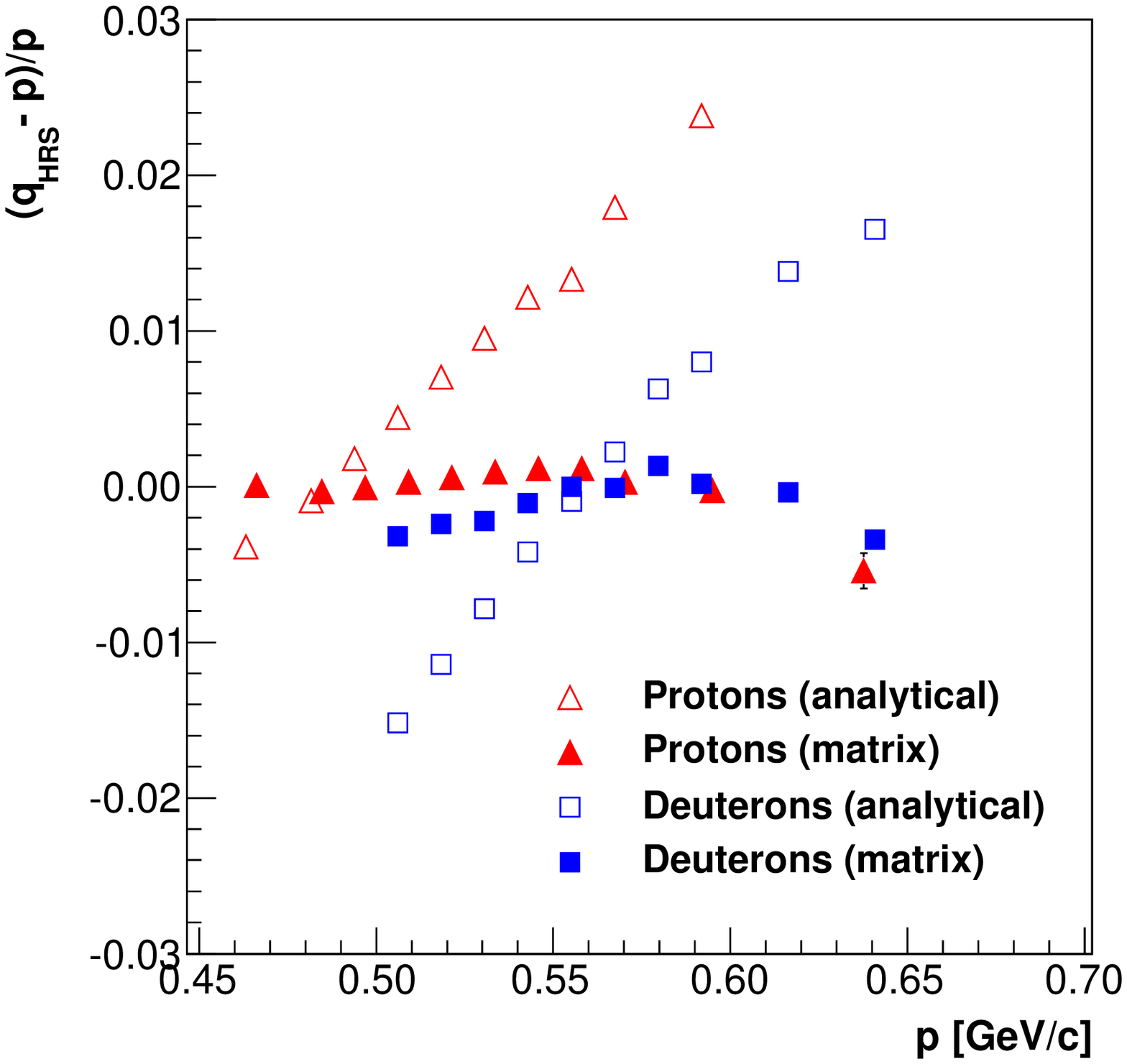}
\caption{[Left] The reconstructed mass of the undetected neutron
(missing mass) from the process $\mathrm{{}^2H(e,e'p)n}$
by using the analytical model and the matrix-formalism (SVD) approach.
The width (sigma) of the peak determined with the analytical model
is $3.3\,\mathrm{MeV}/c^2$ (corresponding to $0.35\,\mathrm{\%}$
relative resolution).  The width of the peak reconstructed
by the SVD method is $4\,\mathrm{MeV}/c^2$.
[Right] The absolute calibration of $\delta_{\mathrm{Tg}}$ as
a function of the particle momentum measured by BigBite.
The relative resolution of $\delta_{\mathrm{Tg}}$
is better in the analytical model than in the matrix method,
but the absolute momentum calibration is inferior to the
matrix approach, except in the narrow region around
$p \approx 0.55\,\mathrm{GeV}/c$.
 \label{MissingMassPlot}}
\end{center}
\end{figure}

With the analytical approximation, resolutions of
a few percent can be achieved, but they deteriorate
when moving towards the edges of the acceptance where
the fringe fields begin to affect the optics.
This is particularly true for $\phi_{\mathrm{Tg}}$.
Figure~\ref{MissingMassPlot} (left) shows the reconstructed mass of 
the neutron from the process $\mathrm{{}^2H(e,e'p)n}$, obtained by using
the analytical model.  The relative resolution is $0.35\,\mathrm{\%}$.
%The $3.7\,\mathrm{MeV}/c^2$ offset is caused by the errors 
%in the energy loss calculation and inherent limitations
%of the model. 

The analytical method requires just a few geometry parameters,
but these need to be known quite accurately.  Had no survey been
performed, the sizes of spectrometer components and the distances
between them could be obtained, in principle, by calibrating
with elastic events.  However, the solution is not unique.
Different combinations of parameters have been shown to yield
almost identical results for the target variables, while
only one combination is correct.

\subsection{The matrix formalism}

In spite of its shortcomings, the analytical model 
is a good starting point.  Due to its simplicity,
it can be implemented and tested quickly, and lends itself well
to online estimation of the experimental data.
For the off-line analysis, a more sophisticated approach
based on the transport matrix formalism is needed.  In this approach,
a prescription is obtained that transforms the detector variables
directly to the target variables.  Various parameterizations
of this transformation are possible.  We have adopted
a polynomial expansion of the form~\cite{bertozzi,nilangaTN}
%\begin{eqnarray}
%  \Omega_{\mathrm{Tg}} = \sum_{i,j,k} \theta_{\mathrm{Det}}^i 
%     y_{\mathrm{Det}}^j \phi_{\mathrm{Det}}^k 
%     \sum_{l} a_{ijkl}^{\Omega_{\mathrm{Tg}}} 
%     x_\mathrm{Det}^l \>,\qquad 
%  \Omega_{\mathrm{Tg}} \in \left\{
%     \delta_{\mathrm{Tg}},\theta_{\mathrm{Tg}},\phi_{\mathrm{Tg}}
%    , y_{\mathrm{Tg}} \right\} \>. \label{eq1}
%\end{eqnarray}
\begin{eqnarray}
  \Omega_{\mathrm{Tg}} = \sum_{i,j,k,l} 
     a_{ijkl}^{\Omega_{\mathrm{Tg}}} \,\,
     x_\mathrm{Det}^i \,
     \theta_{\mathrm{Det}}^j \,
     y_{\mathrm{Det}}^k \,
     \phi_{\mathrm{Det}}^l \>, \qquad 
  \Omega_{\mathrm{Tg}} \in \left\{
     \delta_{\mathrm{Tg}}, \theta_{\mathrm{Tg}},
     \phi_{\mathrm{Tg}}, y_{\mathrm{Tg}} \right\} \>. \label{eq1}
\end{eqnarray}
%This expansion can be interpreted
%as a scalar product of the vector $\{x_{\mathrm{Det}}^l\}$ with
%the vector $\{\theta_{\mathrm{Det}}^i y_{\mathrm{Det}}^j\phi_{\mathrm{Det}}^k\}%$
%and a tensor-like form $a_{ijkl}^{\Omega_{\mathrm{Tg}}}$
%(the optical ``matrix'') containing the expansion parameters.
%Knowing the optics of the spectrometer is equivalent to
%determining the parameters $a_{ijkl}$ and establishing
%the limitations of such a parameterization. 
Knowing the optics of a spectrometer is equivalent to
determining the expansion coefficients $a_{ijkl}^{\Omega_{\mathrm{Tg}}}$ 
(the so-called optical ``matrix'') and establishing
the limitations of such a parameterization. 

Ideally, one would like to obtain a single optical matrix
with full reconstruction functionality for all particle species
and momenta, with as few high-order terms as possible.
In a large-acceptance spectrometer like BigBite,
this represents a considerable challenge.  In particular,
one must clearly understand the contributions of the high-order elements.
Uncontrolled inclusion of these terms typically causes oscillations
of the reconstructed variables at the edges of the acceptance.
In the following sections, we discuss the procedure of constructing
the optical matrix in which special attention is devoted
to checking the convergence of the method and estimating
the robustness of the matrix elements.

\subsubsection{Decoupled description}
\label{sub:direct}

The determination of the optical matrix starts with
a low-order analysis in order to estimate the dominant
matrix elements.  As in the analytical model, the BigBite magnet
is assumed to be an ideal dipole.  This assumption
decouples the in-plane and out-of-plane variables, resulting in
the simplification that $\delta_{\mathrm{Tg}}$ and $\theta_{\mathrm{Tg}}$
depend only on $x_{\mathrm{Det}}$ and $\theta_{\mathrm{Det}}$,
while $y_{\mathrm{Tg}}$ and $\phi_{\mathrm{Tg}}$ depend
only on $y_{\mathrm{Det}}$ and $\phi_{\mathrm{Det}}$.

Since each target coordinate depends only on two detector
coordinates, the matrix elements were estimated by
examining two-dimensional histograms of target coordinates
(as given by the HRS) versus BigBite detector variables,
using various detector-variable cuts.  Since BigBite
in this approximation does not bend horizontally,
only first-order polynomials were utilized to fit the data
for $y_{\mathrm{Tg}}$ and $\phi_{\mathrm{Tg}}$, while expansions
up to third-order were applied for $\delta_{\mathrm{Tg}}$
and $\theta_{\mathrm{Tg}}$:
%\begin{eqnarray}
%\delta_{\mathrm{Tg}}(x_\mathrm{Det}, \theta_\mathrm{Det}) &=& \left[a_{0000}^{\delta_{\mathrm{Tg}}} 
%       +a_{1000}^{\delta_{\mathrm{Tg}}}x_\mathrm{Det} + 
%        a_{2000}^{\delta_{\mathrm{Tg}}}x_\mathrm{Det}^2 \right] \\
%       &+&\left[a_{0100}^{\delta_{\mathrm{Tg}}} +
%        a_{1100}^{\delta_{\mathrm{Tg}}}x_\mathrm{Det} + 
%        a_{2100}^{\delta_{\mathrm{Tg}}}x_\mathrm{Det}^2 \right]\theta_\mathrm{Det} \nonumber\\
%      &+& \left[a_{0200}^{\delta_{\mathrm{Tg}}} 
%       + a_{1200}^{\delta_{\mathrm{Tg}}}x_\mathrm{Det} \right] \theta_\mathrm{Det}^2
%       + \left[a_{0300}^{\delta_{\mathrm{Tg}}} 
%       + a_{1300}^{\delta_{\mathrm{Tg}}}x_\mathrm{Det} \right ]\theta_\mathrm{Det}^3\>, \nonumber  \\
%\theta_{\mathrm{Tg}}(x_\mathrm{Det}, \theta_\mathrm{Det}) &=& \left[a_{0000}^{\theta_{\mathrm{Tg}}} 
%       + a_{1000}^{\theta_{\mathrm{Tg}}}x_\mathrm{Det} + a_{2000}^{\theta_{\mathrm{Tg}}}x_\mathrm{Det}^2 \right] \\
%       &+& \left[a_{0100}^{\theta_{\mathrm{Tg}}} + a_{1100}^{\theta_{\mathrm{Tg}}}x_\mathrm{Det} 
%       + a_{2100}^{\theta_{\mathrm{Tg}}}x_\mathrm{Det}^2 \right] \theta_\mathrm{Det}\>, \nonumber  \\ 
%\phi_{\mathrm{Tg}}(y_\mathrm{Det},\phi_\mathrm{Det}) &=& a_{0000}^{\phi_{\mathrm{Tg}}} 
%       + a_{0001}^{\phi_{\mathrm{Tg}}}\phi_\mathrm{Det}\>, \nonumber \\ 
%y_{\mathrm{Tg}}(y_\mathrm{Det}, \phi_\mathrm{Det}) &=& \left[a_{0001}^{y_{\mathrm{Tg}}}
%       + a_{0011}^{y_{\mathrm{Tg}}}y_\mathrm{Det} \right]\phi_\mathrm{Det}  + \left[ a_{0000}^{y_{\mathrm{Tg}}} 
%       + a_{0010}^{y_{\mathrm{Tg}}}y_\mathrm{Det}\right]\,. \nonumber 
%\end{eqnarray}
\begin{eqnarray}
\delta_{\mathrm{Tg}}(x, \theta) &=& \left[a_{0000}^{\delta_{\mathrm{Tg}}} 
       +a_{1000}^{\delta_{\mathrm{Tg}}}x + 
        a_{2000}^{\delta_{\mathrm{Tg}}}x^2 \right]
       +\left[a_{0100}^{\delta_{\mathrm{Tg}}} +
        a_{1100}^{\delta_{\mathrm{Tg}}}x + 
        a_{2100}^{\delta_{\mathrm{Tg}}}x^2 \right]\theta \nonumber\\
      &+& \left[a_{0200}^{\delta_{\mathrm{Tg}}} 
       + a_{1200}^{\delta_{\mathrm{Tg}}}x \right] \theta^2
       + \left[a_{0300}^{\delta_{\mathrm{Tg}}} 
       + a_{1300}^{\delta_{\mathrm{Tg}}}x \right ]\theta^3\>, \nonumber  \\
\theta_{\mathrm{Tg}}(x, \theta) &=& \left[a_{0000}^{\theta_{\mathrm{Tg}}} 
       + a_{1000}^{\theta_{\mathrm{Tg}}}x + a_{2000}^{\theta_{\mathrm{Tg}}}x^2 \right]
       + \left[a_{0100}^{\theta_{\mathrm{Tg}}} + a_{1100}^{\theta_{\mathrm{Tg}}}x 
       + a_{2100}^{\theta_{\mathrm{Tg}}}x^2 \right] \theta\>, \nonumber  \\ 
\phi_{\mathrm{Tg}}(y,\phi) &=& a_{0000}^{\phi_{\mathrm{Tg}}} 
       + a_{0001}^{\phi_{\mathrm{Tg}}}\phi\>, \nonumber \\ 
y_{\mathrm{Tg}}(y, \phi) &=& \left[a_{0001}^{y_{\mathrm{Tg}}}
       + a_{0011}^{y_{\mathrm{Tg}}}y \right]\phi  + \left[ a_{0000}^{y_{\mathrm{Tg}}} 
       + a_{0010}^{y_{\mathrm{Tg}}}y\right]\,. \nonumber 
\end{eqnarray}
The calculated matrix elements are shown in the second column
of Table~\ref{table1}.  The $a_{0001}^{\phi_{\mathrm{Tg}}}$ matrix element 
was set to $1$ since there is no in-plane bending.
This approximation could not be used for further physics analysis
because higher-order corrections are needed.  However,
the low-order terms are very robust and do not change much when
more sophisticated models with higher-order terms are considered.
The results obtained by using this method serve
as a benchmark for more advanced methods, in particular
as a check whether the matrix elements computed by automated
numerical algorithms converge to reasonable values.

\begin{table}[!ht]
\caption{The dominant matrix elements of the BigBite
optics model (Eq.~(\ref{eq1})) determined by a decoupled
description (subsubsection~\ref{sub:direct}),
by simplex minimization (N\&M), and by singular
value decomposition (SVD, subsubsection~\ref{sub:nmsvd}).
\label{table1}}
\begin{center}
% use packages: array
\begin{tabular}{lrrr}
\hline

\hline

\hline
 Matrix & Decoupled & N\&M & SVD \\[-6pt]
 element & description & & \\ 
\hline
$a_{0010}^{y_\mathrm{Tg}}\,[\mathrm{m/m}]$ 
  & $0.998$ & $1.024$ & $0.917$ \\
$a_{0001}^{y_\mathrm{Tg}}\,[\mathrm{m/rad}]$ 
  & $-2.801$ & $-2.839$ & $-2.766$ \\ 
\hline
$a_{0001}^{\phi_\mathrm{Tg}}\,[\mathrm{rad/rad}]$ 
  & $1.000$ & $1.052$ &  $0.9517$ \\ 
\hline
$a_{1000}^{\theta_\mathrm{Tg}}\,[\mathrm{rad/m}]$ 
  & $0.497$ & $0.549$ & $0.551$ \\
$a_{0100}^{\theta_\mathrm{Tg}}\,[\mathrm{rad/rad}]$ 
  & $-0.491$ & $-0.490$ & $-0.484$ \\
\hline
$a_{1000}^{\delta_\mathrm{Tg}}\,[\mathrm{1/m}]$ 
  & $-0.754$ & $-0.716$ & $-0.676$ \\
$a_{0100}^{\delta_\mathrm{Tg}}\,[\mathrm{1/rad}]$ 
  & $2.811$ & $2.881$ & $2.802$ \\ 
\hline

\hline

\hline
%\multicolumn{5}{l}{\begin{scriptsize}${}^\dagger$ $a_{0001}^{\phi_{Tg}}$ was a priori set to $1$ in this approximation. \end{scriptsize}}
\end{tabular}
\end{center}
\end{table}

\subsubsection{Higher order matrix formalism}
\label{sub:nmsvd}

For the determination of the optics matrix a numerical method
was developed in which matrix elements up to fourth order
were retained.  Their values were calculated by using
a $\chi^2$-minimization scheme, wherein the target
variables calculated by Eq.~(\ref{eq1}) were compared
to the directly measured values,
\begin{eqnarray}
  \chi^2\left(a_i^{\Omega_{\mathrm{Tg}}}\right) = 
     \sqrt{\left( \Omega_{\mathrm{Tg}}^{\mathrm{Measured}} -
     \Omega_{\mathrm{Tg}}^{\mathrm{Optics}}\left(x_{\mathrm{Det}},
      y_{\mathrm{Det}}, \theta_{\mathrm{Det}}, \phi_{\mathrm{Det}}; 
      a_i^{\Omega_{\mathrm{Tg}}}\right) \right)^2} \>, \qquad
      i = 1,2,\,\ldots\>,M\,. \label{eq2}
\end{eqnarray}
The use of $M$ matrix elements for each target variable means
that a global minimum in $M$-dimensional space must be found.
Numerically this is a very complex problem; two techniques
were considered for its solution.

Our first choice was the downhill simplex method developed
by Nelder and Mead \cite{nelder,nrc}.  The method tries
to minimize a scalar non-linear function of $M$ parameters
by using only function evaluations (no derivatives).
It is widely used for non-linear unconstrained optimization,
but it is inefficient and its convergence properties
are poorly understood, especially in multi-dimensional minimizations.
The method may stop in one of the local minima instead of the global
minimum \cite{lagarias,mckinnon}, so an additional examination
of the robustness of the method was required.

The set of functions $\Omega_{\mathrm{Tg}}$ is linear
in the parameters $a_i^{\Omega_{\mathrm{Tg}}}$.
Therefore, Eq.~(\ref{eq2}) can be written as
\begin{eqnarray}
  \chi^2 = \sqrt{ \, \left| A \, \vec{a} - \vec{b} \, \right |^2}\>,\label{eq3}
\end{eqnarray}
where the $M$-dimensional vector $\vec{a}$ contains
the matrix elements $a_i^{\Omega_{\mathrm{Tg}}}$,
and the $N$-dimensional vector $\vec{b}$ contains
the measured values of the target variable being considered.
The elements of the $N\times M$ matrix $A$ are various products
of detector variables ($x_\mathrm{Det}^i \theta_{\mathrm{Det}}^j
y_{\mathrm{Det}}^k \phi_{\mathrm{Det}}^l$) for each measured event.
The system $A\,\vec{a} = \vec{b}$ in Eq.~(\ref{eq3})
is overdetermined ($N>M$), thus the vector $\vec{a}$
that minimizes the $\chi^2$ can be computed by singular
value decomposition (SVD).  It is given by $A = UWV^\mathrm{T}$,
where $U$ is a $N\times M$ column-orthogonal matrix,
$W$ is a $M\times M$ diagonal matrix with non-negative
singular values $w_i$ on its diagonal, and $V$ is a $M\times M$
orthogonal matrix \cite{golub,nrc}.  The solution has the form
$$
\vec{a} = \sum_{i=1}^M\left(\frac{ \vec{U}_i\cdot\vec{b}}{w_i} \right)
    \vec{V}_i \>.
$$

The SVD was adopted as an alternative to simplex minimization
since it produces the best solution in the least-square sense,
obviating the need for robustness tests.  Another great advantage
of SVD is that it can not fail; the method always returns a solution,
but its meaningfulness depends on the quality of the input data.
The most important leading-order matrix elements computed
by using both techniques are compared in Table~\ref{table1}.

\section{Calibration results}

\subsection{Vertex position}

\begin{figure}[!ht]
\begin{center}
\includegraphics[width=0.49\textwidth]{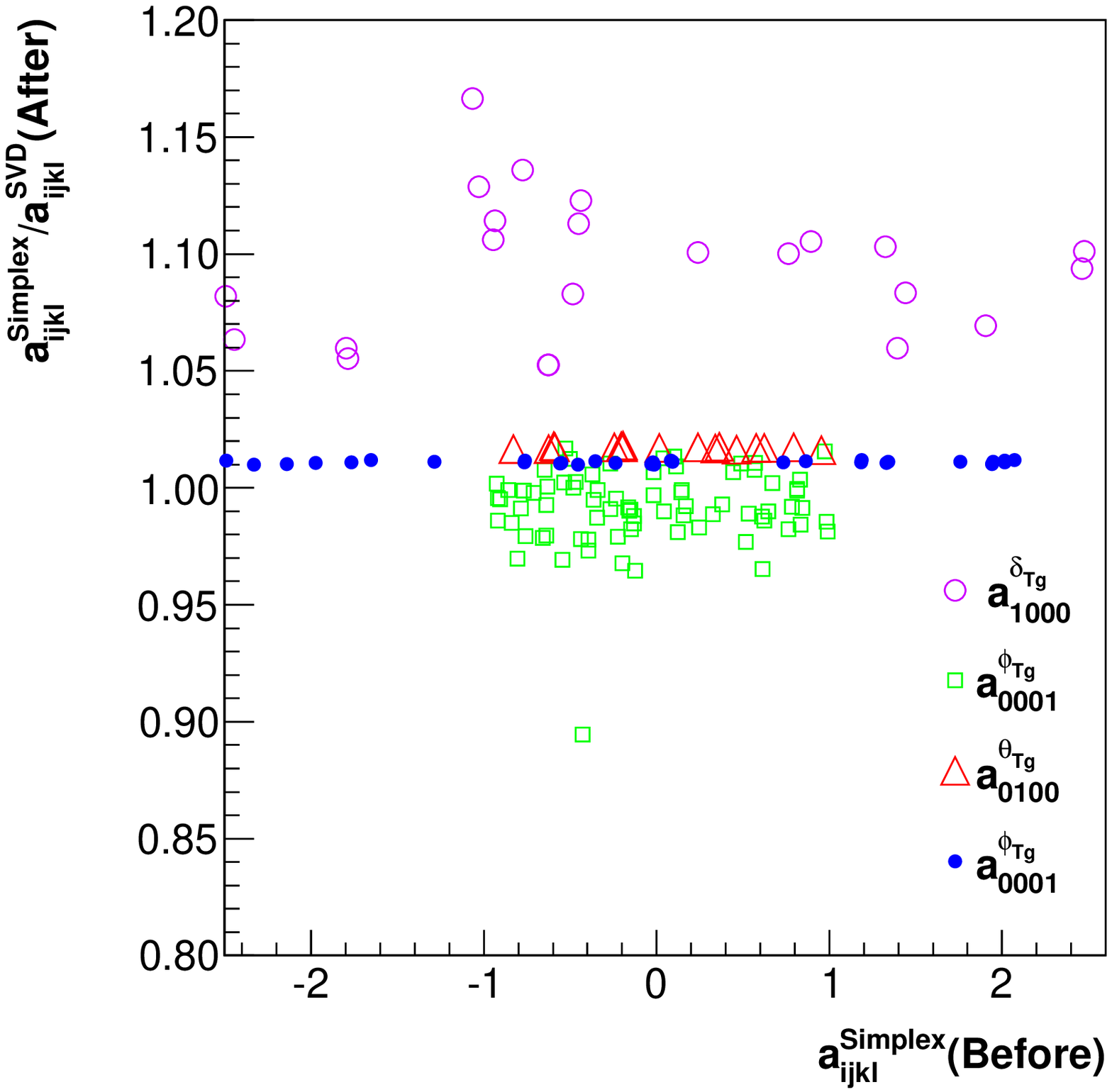}
\includegraphics[width=0.49\textwidth]{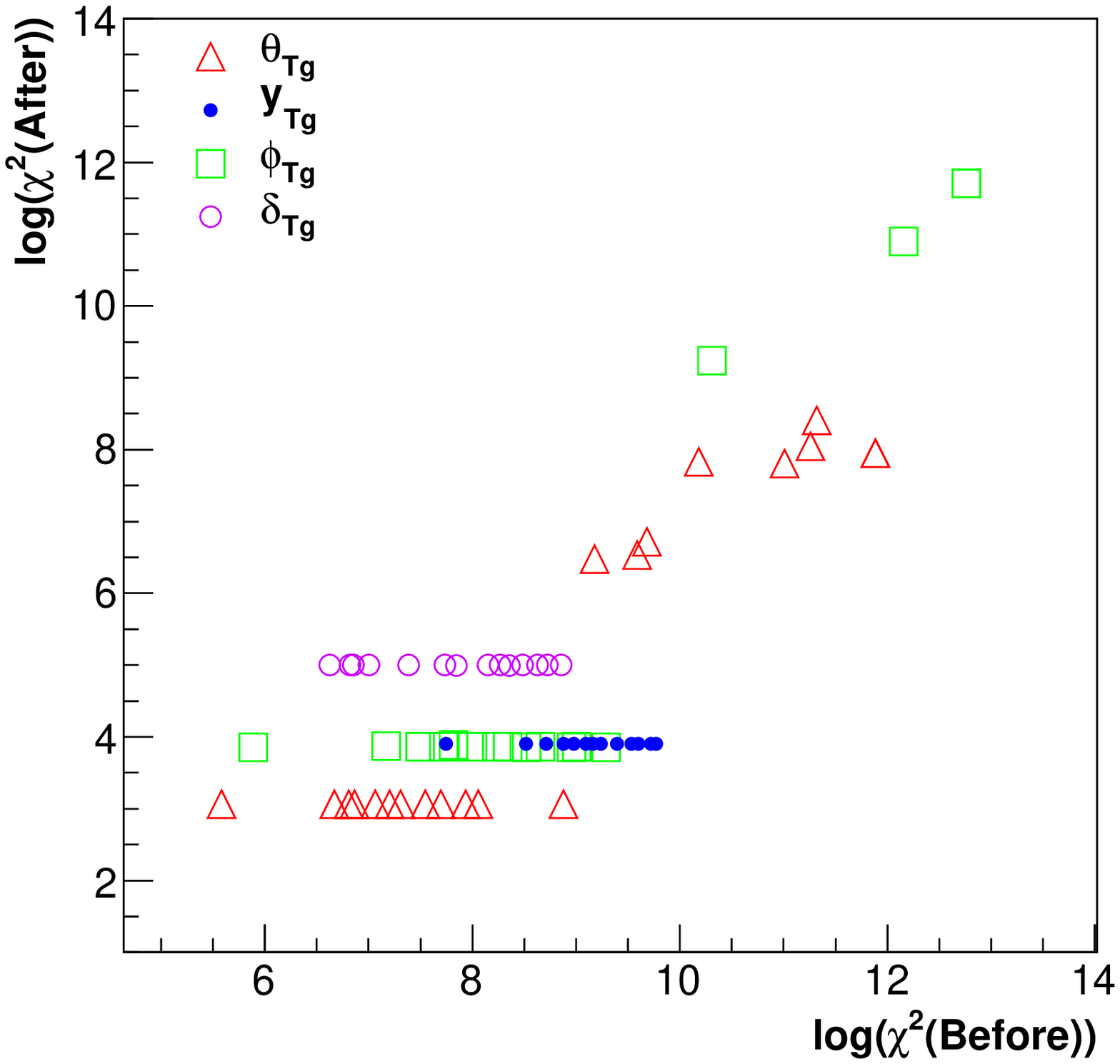}
\caption{[Left] Robustness checks of the simplex minimization method
for select matrix elements  $a_{ijkl}^{\Omega_{\mathrm{Tg}}}$.
The analysis was done for a large set of randomly chosen initial conditions
for each target coordinate.  The fact that the vast majority
of the initial conditions converge to a single value
is an indication of the robustness of the method.
[Right] The values of the $\chi^2$-function before and after
simplex minimization for all four target coordinates.
The method converges to a single $\chi^2$ value
for a wide range of initial conditions (note the log scales).
The solution with the smallest $\chi^2$ represents the result
used in the optics-matrix. \label{TargetYConvergence} }
\end{center}
\end{figure}

The matrix for the vertex position variable $y_{\mathrm{Tg}}$
was obtained by analyzing the protons from quasi-elastic scattering
of electrons on the multi-foil carbon target.  The positions
of the foils were measured by a geodetic survey to sub-millimeter
accuracy, allowing for a very precise calibration of $y_{\mathrm{Tg}}$.
The vertex information from the HRS was used to locate
the foil in which the particle detected by BigBite originated.
This allowed us to directly correlate the detector variables
for each coincidence event to the interaction vertex.
When Eq.~(\ref{eq1}) is written for $y_{\mathrm{Tg}}$,
a linear equation for each event can be formed:
\begin{eqnarray}
 {y_\mathrm{Tg}}_{(n)}^{\mathrm{Measured}} = 
   {y_{\mathrm{Tg}}}_{(n)}^{\mathrm{Optics}} 
   &:=& a_{0000}^{y} + a_{0001}^{y}\phi_{(n)} + 
       a_{0002}^{y}\phi_{(n)}^2 + a_{0003}^{y}\phi_{(n)}^3 + 
       \cdots\ \nonumber\\
   &+& a_{0010}^{y}y_{(n)} + a_{0020}^{y}y_{(n)}^2 + 
       a_{0030}^{y}y_{(n)}^3 + a_{0040}^{y}y_{(n)}^4 + 
       \cdots\ \nonumber \\
   &+& a_{0100}^{y}\theta_{(n)} + a_{0200}^{y}\theta_{(n)}^2+ 
       a_{0300}^{y}\theta_{(n)}^3 + a_{0400}^{y}\theta_{(n)}^4 + 
       \cdots \nonumber \\
   &+& a_{1000}^{y}x_{(n)} + a_{2000}^{y}x_{(n)}^2 + 
       a_{3000}^{y}x_{(n)}^3 + a_{4000}^{y}x_{(n)}^4 + 
       \cdots\ \nonumber \\
   &+&   a_{1111}^{y}x_{(n)}\theta_{(n)} y_{(n)} \phi_{(n)}\>, 
        \label{TgYAnsatz}
\end{eqnarray}
where $n = 1,2,\ldots,N$, and $N$ is the number of coincidence events
used in the analysis. The overdetermined set of Eqs.~(\ref{TgYAnsatz})
represents a direct comparison of the reconstructed vertex position
$y_{\mathrm{Tg}}^{\mathrm{Optics}}$ to the measured value
$y_{\mathrm{Tg}}^{\mathrm{Measured}}$.  Initially a consistent
polynomial expansion to fourth degree ($i+j+k+l \leq 4$)
was considered, which depends on $70$ matrix elements $a_{ijkl}^{y}$.
Using this ansatz in Eq.~(\ref{eq2}) defines a $\chi^2$-minimization
function, which serves as an input to the simplex method.
To be certain that the minimization did not converge to one of
the local minima, the robustness of this method was examined
by checking the convergence of the minimization algorithm
for a large number of randomly chosen initial sets of parameters 
(see Fig.~\ref{TargetYConvergence}).

The results were considered to be stable if the $\chi^2$
defined by Eq.~(\ref{eq2}) converged to the same value
for the majority of initial conditions.
Small variations in $\chi^2$ were allowed: they are caused
by small matrix elements which are irrelevant for $y_{\mathrm{Tg}}$,
but have been set to non-zero values in order to additionally minimize $\chi^2$
in a particular minimization process.  These matrix elements
could be easily identified and excluded during the robustness checks
because they are unstable and converge to a different value
in each minimization. Ultimately only $25$ matrix elements that had the 
smallest fluctuations were kept for the $y_{\mathrm{Tg}}$ matrix.

The SVD method was used next.  To compute the matrix elements
for $y_{\mathrm{Tg}}$, the linear set of Eqs.~(\ref{TgYAnsatz})
first needs to be rewritten in the form $A\,\vec{a} = \vec{b}$
used in Eq.~(\ref{eq3}):
\begin{eqnarray}
 \left( 
  \begin{array}{cccc}
  1 & \phi_{(1)} & \cdots & x_{(1)}\theta_{(1)}y_{(1)}\phi_{(1)} \\ 
  1 & \phi_{(2)} & \cdots & x_{(2)}\theta_{(2)}y_{(2)}\phi_{(2)} \\ 
  1 & \phi_{(3)} & \cdots & x_{(3)}\theta_{(3)}y_{(3)}\phi_{(3)} \\ 
  \vdots & \vdots  & \ddots & \vdots \\ 
  1 & \phi_{(N-2)} & \cdots & x_{(N-2)}\theta_{(N-2)}y_{(N-2)}\phi_{(N-2)} \\ 
  1 & \phi_{(N-1)} & \cdots & x_{(N-1)}\theta_{(N-1)}y_{(N-1)}\phi_{(N-1)} \\ 
  1 & \phi_{(N)} & \cdots & x_{(N)}\theta_{(N)}y_{(N)}\phi_{(N)} \\ 
  \end{array}
  \right) 
 \left( 
  \begin{array}{c}
  a_{0000} \\ 
  a_{0001} \\ 
  \vdots \\
  a_{1111}
  \end{array}
 \right)  = 
 \left( 
  \begin{array}{c}
  {y_{\mathrm{Tg}}}_{(1)} \\ 
  {y_{\mathrm{Tg}}}_{(2)} \\ 
  {y_{\mathrm{Tg}}}_{(3)} \\ 
  \vdots \\
  {y_{\mathrm{Tg}}}_{(N-2)} \\
  {y_{\mathrm{Tg}}}_{(N-1)} \\
  {y_{\mathrm{Tg}}}_{(N)}
  \end{array}
 \right)\>, \nonumber
\end{eqnarray}
where $\vec{a}$ contains $M$ unknown matrix elements $a_{ijkl}^{y}$
to be determined by the SVD, $\vec{b}$ contains $N$ measured values
of $y_{\mathrm{Tg}}$, and $A$ is filled with the products
of detector variables accompanying the matrix elements
in the polynomial expansion of Eq.~(\ref{TgYAnsatz}) for each event.

\begin{figure}[!ht]
\begin{center}
\includegraphics[width=0.6\textwidth]{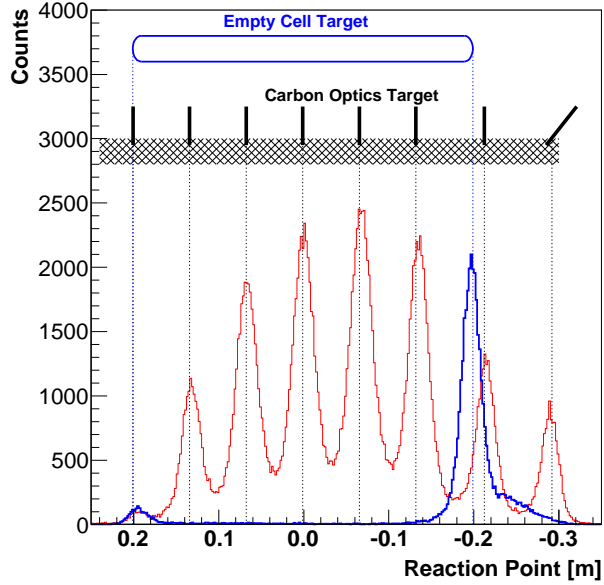}
\caption{The reconstructed vertex position (reaction point)
for the multi-foil carbon target and the empty cell
of the production target, by using the SVD technique.
The vertical dashed lines indicate the actual positions
of the carbon foils and the empty-cell glass windows.
The small shoulder to the right of the reconstructed empty-cell
entry window is due to the jet of $^4\mathrm{He}$ gas used
to cool the window at the beam impact point. \label{TgYResult}}
\end{center}
\end{figure}

The SVD analysis also began with $70$ matrix elements, 
but was not applied to one combined data set as in the 
simplex method in order to extract the most relevant ones.
Rather, it was used on each set of data separately. 
From the comparison of the matrix elements obtained
with different calibration data sets, only the elements 
fluctuating by less than $100\,\%$ were selected. Although
this choice appears to be arbitrary, the results do not change much by 
modifying this criterion, for example, by including elements
with as much as $\pm 1000\, \mathrm{\%}$ fluctuation.
The final set of matrix elements contained only $37$ of the best entries.
With these elements,  the entire analysis was repeated in order
to calculate their final values.  The most relevant elements
are listed in Table~\ref{table1}.  The result of the calibration
of $y_\mathrm{Tg}$ is shown in Fig.~\ref{TgYResult}.

\subsection{Angular coordinates}

For the calibration of the angular variables $\theta_{\mathrm{Tg}}$
and $\phi_{\mathrm{Tg}}$, a set of quasi-elastic data on carbon
and deuterium targets taken with the sieve-slit collimator was analyzed.
The particles that pass through different holes can be well separated
and localized at the detector plane.

\begin{figure}[hbtp]
\begin{center}
\includegraphics[width=0.65\textwidth]{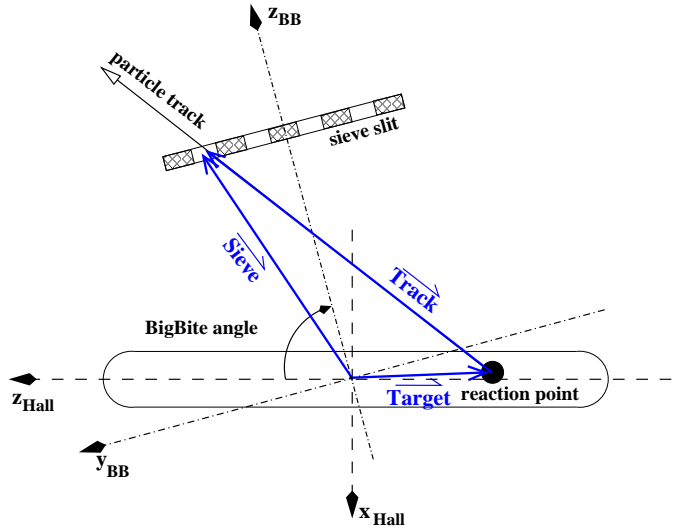} 
\caption{Position of the sieve-slit collimator relative to the target.
The vector of the particle track through a particular hole
in the sieve is the difference of the position vector at the hole
and the reaction-point vector.  BigBite is positioned
at $-75^\circ$ with respect to the beam direction. Other settings are 
listed in Table~\ref{table_kinematics}.
\label{BBSieveSlitDiagram}}
\end{center}
\end{figure}

By knowing the detector coordinates and the accurate position
of the corresponding hole in the sieve, the target variables
can be calculated.  From the reaction point at the target
(see Fig.~\ref{BBSieveSlitDiagram}), $\theta_{\mathrm{Tg}}$
and $\phi_{\mathrm{Tg}}$ can be calculated:
\begin{eqnarray}
  \tan \phi_{\mathrm{Tg}} = \frac{y_{\mathrm{Sieve}} - 
    y_{\mathrm{Tg}}}{z_{\mathrm{Sieve}} - z_{\mathrm{Tg}}} \>,
    \qquad\tan \theta_{\mathrm{Tg}} = 
    \frac{x_{\mathrm{Sieve}} - 
    x_{\mathrm{Tg}}}{z_{\mathrm{Sieve}} - z_{\mathrm{Tg}}} \>. \nonumber
\end{eqnarray}

By using the values of the target variables, a set of linear
equations has been written for all measured events, and matrix
elements determined by using both numerical approaches.
In the simplex method, $30$ matrix elements for $\theta_{\mathrm{Tg}}$
and $68$ elements for $\phi_{\mathrm{Tg}}$ were retained.
Robustness checks for both angular variables were repeated
to ensure that the global minimum had been reached.

\begin{figure}[!ht]
\begin{center}
\includegraphics[width=0.6\textwidth]{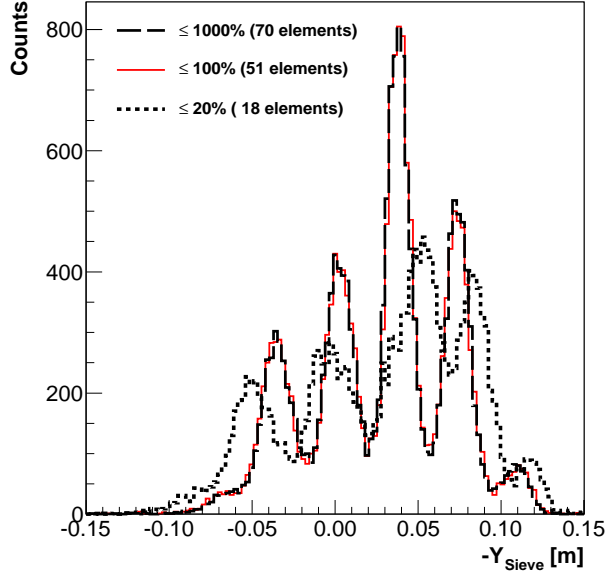}
\caption{The reconstructed positions of the holes in the top row
of the sieve-slit collimator, computed from $\phi_\mathrm{Tg}$.
The quality of the reconstruction depends on the number of included
matrix elements.  There is almost no difference when the elements
fluctuating by up to $\pm 1000\,\%$ are retained ($70$ elements,
dashed lines) or only those that fluctuate by up to $\pm 100\,\%$
($51$ elements, full line).  The quality deteriorates if too many
elements are dropped (i.e.~keeping $18$ elements fluctuating
by less than $\pm 20\,\%$, dotted lines). \label{TgPhME}}
\end{center}
\end{figure}

The SVD analysis also started with $70$ matrix elements,
which were ultimately reduced to $37$ for $\theta_{\mathrm{Tg}}$
and $51$ for $\phi_{\mathrm{Tg}}$, again taking into account
only those elements that fluctuated by less than $100\,\mathrm{\%}$.
Figure~\ref{BBSieve} (right) shows the reconstructed sieve pattern.
The majority of the holes are reconstructed, except those obscured
by parts of the experimental apparatus due to specific geometric
constraints during the experiment.  In order to demonstrate
the effect of gradually excluding redundant matrix elements,
Fig.~\ref{TgPhME} shows the reconstructed top row of the sieve-slit
collimator holes when the elements with up to $\pm 1000\,\%$,
$\pm 100\,\%$, and $\pm 20\,\%$ fluctuations are retained.
There is virtually no difference in the reconstructed pattern
when all elements exceeding the $\pm 100\,\%$ fluctuations are dropped,
while errors start to appear when those fluctuating by less than
$\pm 100\,\%$ are dropped.

\begin{figure}[hbtp]
\begin{center}
\includegraphics[width=0.48\textwidth]{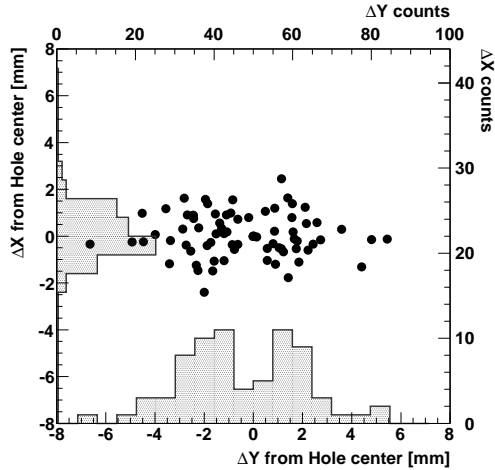}
\caption{Distribution of vertical ($\Delta X$) and horizontal ($\Delta Y$)
deviation of the center of each reconstructed sieve-slit hole
from its true position. Observed 
deviations are much smaller than the diameter of a hole, which is  
$19.1\,\mathrm{mm}$.  The horizontal and vertical histograms (top and right
axis labels, respectively) represent the distributions
in the horizontal and vertical directions.
\label{ThPhCenterPositionError}}
\end{center}
\end{figure}

The quality of the sieve-pattern
reconstruction was examined by comparing the centers of the reconstructed
holes with their true positions.  Figure~\ref{ThPhCenterPositionError}
shows that, with the exception of a few holes near the acceptance edges,
these deviations are smaller than $2\,\mathrm{mm}$ in the vertical,
and smaller than $4\,\mathrm{mm}$ in the horizontal direction.
This is much less than the hole diameter, which is  $19.1\,\mathrm{mm}$.

Once the sieve pattern was reconstructed, an absolute calibration
had to be performed to correct for any BigBite misalignment
and mispointing.  For that purpose hydrogen and deuterium elastic data 
were used. By comparing the direction of the momentum transfer vector
from the HRS to the calculated values of $\theta_{\mathrm{Tg}}$
and $\phi_{\mathrm{Tg}}$, the zero-order matrix elements could be
properly determined and the offsets corrected.  In addition,
the precise distance between the target and the sieve-slit collimator
was obtained, which we were not able to measure precisely
due to physical obstacles between the target and BigBite.
From this analysis, the sieve slit was determined to be 
positioned $1.13\,\mathrm{m}$ away from the target.

\subsection{Momentum}

\begin{figure}[hbtp]
\begin{center}
\includegraphics[width=0.49\textwidth]{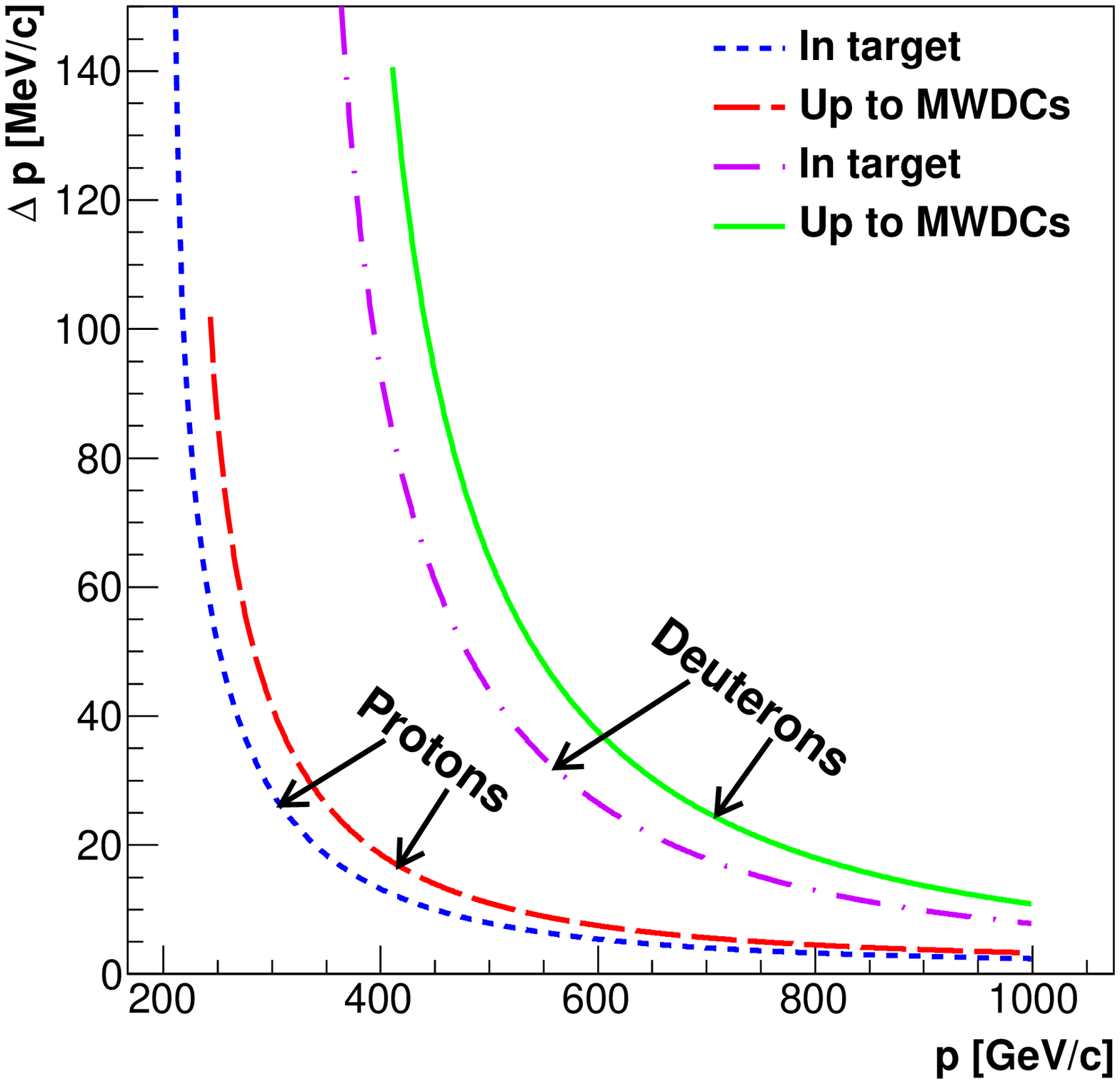}
\includegraphics[width=0.49\textwidth]{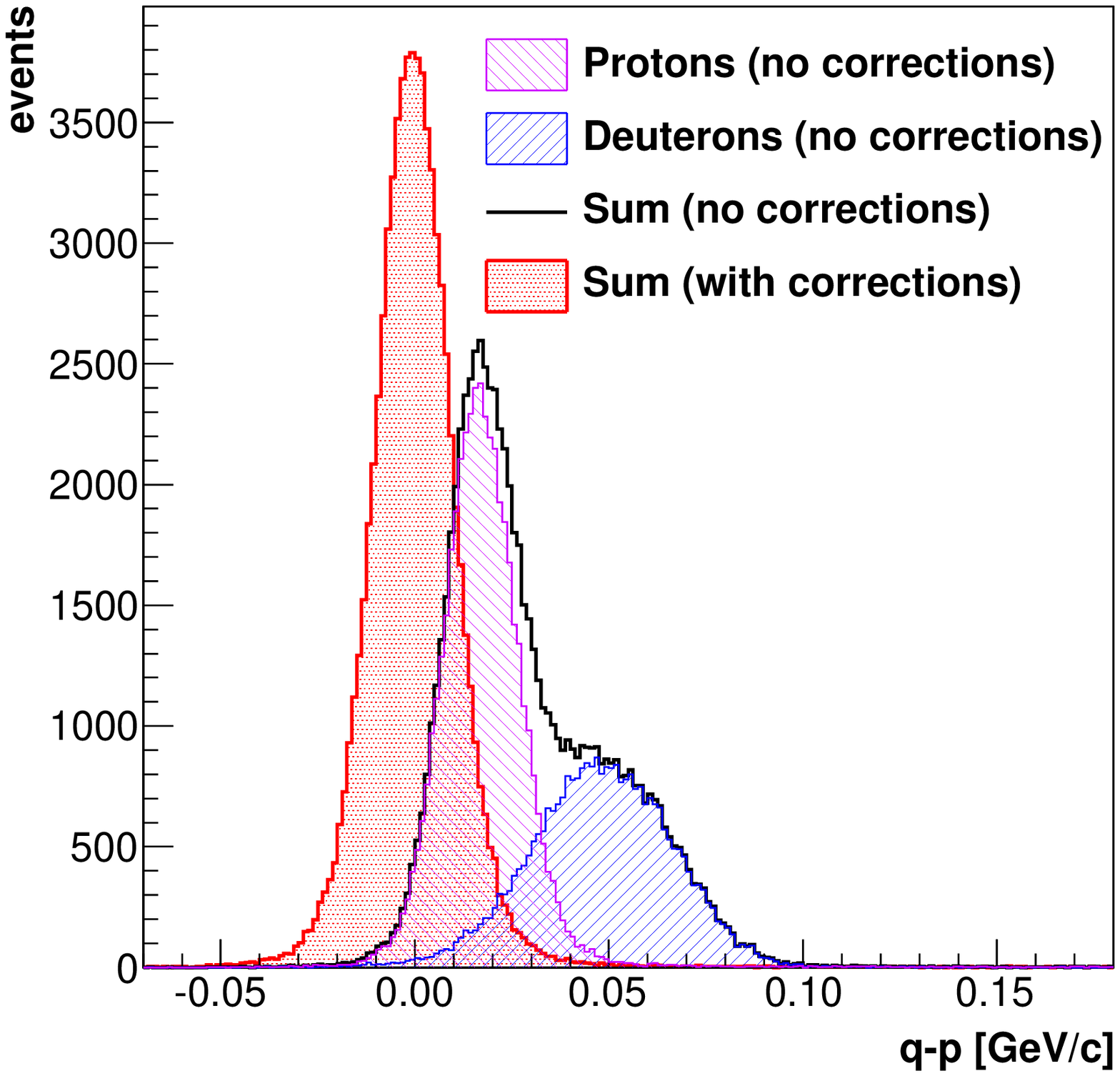}
\caption{[Left] Momentum losses of protons and deuterons
inside the target and the total momentum losses up to the MWDCs.
[Right] Quality of reconstructed momentum for elastic protons and deuterons. 
If energy losses are not taken into account, two peaks are visible (center
and right histograms summed to the full curve). With proper inclusion of 
energy losses both peaks merge into one (left histogram), resulting in
better momentum resolution.  \label{figure_MomentumLosses}}
\end{center}
\end{figure}

The matrix elements for the $\delta_{\mathrm{Tg}}$ variable
were obtained by using data from elastic scattering of electrons
on hydrogen and deuterium for which the particle momentum
in BigBite should be exactly the same as the momentum 
transfer $\vec{q}$ given by the HRS.  We assumed that 
$\delta_{\mathrm{Tg}}$ depends only on $x_{\mathrm{Det}}$ and 
$\theta_{\mathrm{Det}}$, while the dependencies involving 
$y_{\mathrm{Det}}$ and $\phi_{\mathrm{Det}}$ were neglected.
Furthermore, the use of in-plane coordinates in the analysis
for $\delta_{\mathrm{Tg}}$ could result in an erroneous matrix due to 
the strong $\phi_{\mathrm{Tg}}$ dependence inherent
to elastic scattering (events strongly concentrated at one edge
of the acceptance).  Considering only $x_{\mathrm{Det}}$ and
$\theta_{\mathrm{Det}}$ matrix elements, $\delta_{\mathrm{Tg}}$ 
can be expressed as
\begin{eqnarray}
  \delta_{\mathrm{Tg}} = \frac{q_{\mathrm{HRS}} - \Delta_{\mathrm{Loss}}}
   {p_\mathrm{c}}-1 = 
   a_{0000}^\delta + a_{1000}^\delta x_{\mathrm{Det}} + 
   a_{0100}^\delta\theta_{\mathrm{Det}}+\cdots \,. 
   \label{momentumeq}
\end{eqnarray}
In order to obtain the optics matrix applicable to all types
of particles, energy losses $\Delta_{\mathrm{Loss}}$ for particle 
transport through the target enclosure and materials within the 
BigBite spectrometer were studied carefully.  
The energy losses were estimated by the Bethe-Bloch formula \cite{leo},
but since the losses were significant, the formula had to be
integrated over the complete particle track for each particle
type and each initial momentum.  The two largest contributions
to the total momentum loss came from the target cell walls
and from the air between the target and the detectors.
(The latter losses could be alleviated by using a helium bag
between the target and the detectors, but its benefits were
considered to be smaller than the technical problems involved.)
The resulting corrections that were taken into account in
Eq.~(\ref{momentumeq}) are shown in Fig.~\ref{figure_MomentumLosses} (left). 

The elastic data available for calibration (momentum range
approximately $0.45\,\mathrm{GeV}/c$ to $0.7\,\mathrm{GeV}/c$)
covered only about half of the BigBite momentum acceptance.
To calibrate the low-momentum region from $0.2\,\mathrm{GeV}/c$
to $0.45\,\mathrm{GeV}/c$, we used protons from quasi-elastic 
scattering on ${}^3\mathrm{He}$ by exploiting the information
from the scintillator dE- and E-planes; the deposited particle
energy in each plane was directly mapped to the particle momentum,
based on known properties of the scintillator material.
The punch-through point, corresponding to the particular
momentum at which the particle has just enough energy
to penetrate through the scintillators, served as a reference.

\begin{figure}[!th]
\begin{center}
\includegraphics[width=0.49\textwidth]{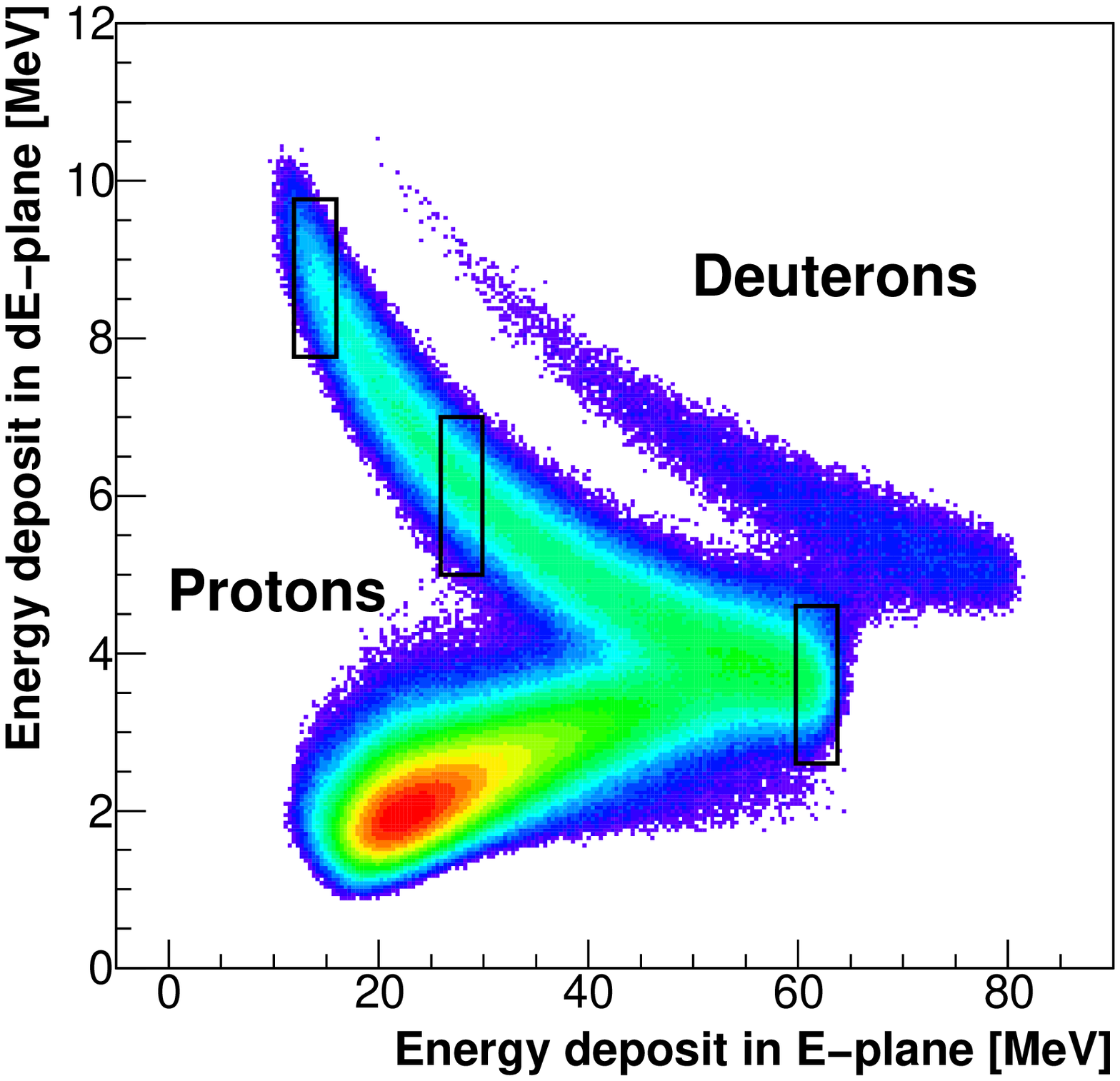}
\includegraphics[width=0.49\textwidth]{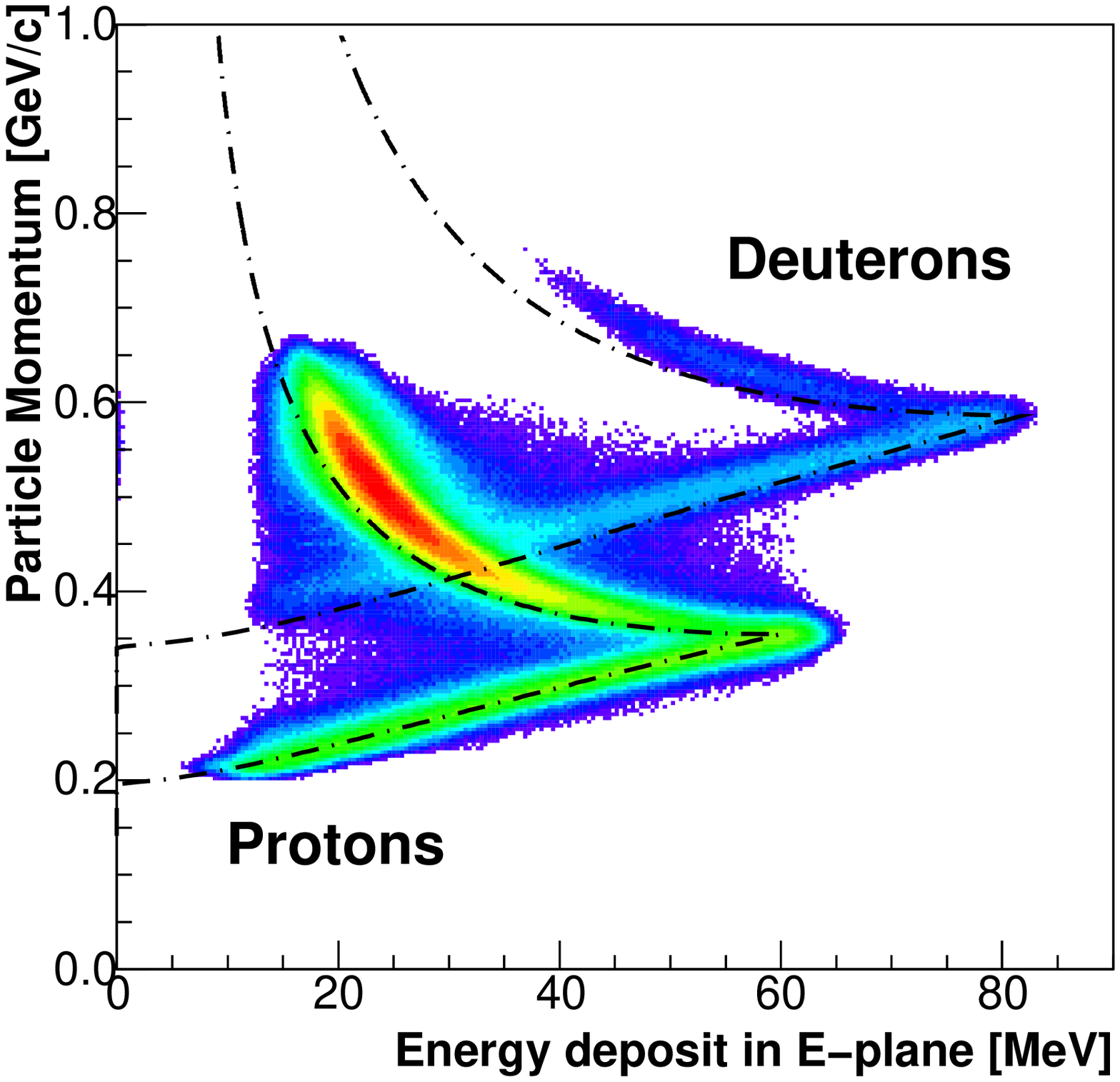}
\caption{[Left] The energy losses in the thin ($3\,\mathrm{mm}$)
scintillator dE-plane versus the energy losses in the thicker
($3\,\mathrm{cm}$) E-plane.  The punch-through points, at which
the protons and deuterons have just enough energy to penetrate
both scintillation planes, are clearly visible.  The black boxes
show sections of events with precisely determined momenta
that were used in the $\delta_{\mathrm{Tg}}$ calibration.
[Right] Particle momentum as a function of energy losses in the E-plane
for ${}^3\mathrm{He}$ data.  The deuterons can be clearly distinguished
from the protons.  The measurements agree well with the simulation
(dot-dashed line).\label{figure_EdE}}
\end{center}
\end{figure}

Beside the proton punch-through point, two other points
with exactly known energy deposits in the dE- and E-planes
were identified, as illustrated in Fig.~\ref{figure_EdE}.  With the 
additional information from these points, a complete momentum
calibration was possible.  To compute the $\delta_{\mathrm{Tg}}$
matrix elements, both numerical approaches described above
were used.  Since the available data were rather sparse,
the search for the most stable matrix elements was not performed
and a complete expansion to fifth order was considered
in both techniques.  Since only a two-variable dependency
was assumed, a complete description was achieved 
by using only $21$ matrix elements.

The comparison of the most relevant matrix elements
obtained from both numerical approaches is again shown
in Table~\ref{table1}.  Figure~\ref{figure_EdE} (right) shows
that the $\delta_{\mathrm{Tg}}$ matrix is well under control.
The reconstructed momentum agrees with the simulation
of energy losses inside the scintillation planes
for the complete momentum acceptance of BigBite,
for both protons and deuterons.  Figure~\ref{MissingMassPlot}
shows the missing-mass peak for the $\mathrm{{}^2H( e, e'p)n}$ process.
The resolution of the reconstructed neutron mass is approximately
$4\,\mathrm{MeV}/c^2$. 

\subsection{Resolution}

The quality of the BigBite optics was also studied.
The resolution of the vertex position was estimated
from the difference between the reconstructed $y_{\mathrm{Tg}}$
and the true position at the target by taking the width
(sigma) of the obtained distribution.  This part of the analysis 
was done by using $2$-pass ($2.425\,\mathrm{GeV}$ beam)
quasi-elastic carbon data.  The extracted values
for the resolution of $y_\mathrm{Tg}$ in different momentum bins
can be parameterized as
$$
\sigma_{y_\mathrm{Tg}} \approx 0.01\left(1 + \frac{0.02}{p^{4}} \right)\>,
$$ 
where the particle momentum is in $\mathrm{GeV}/c$ and the result is in meters.
It is best at the upper limit of the accepted momentum range 
(about $p=0.7\,\mathrm{GeV}/c$) where it amounts to 
$\sigma_{y_{\mathrm{Tg}}}=1.1\,\mathrm{cm}$.
The deterioration of the resolution at lower momenta is due to
multiple scattering~\cite{leo} in the air between the scattering
chamber and the MWDCs.

\begin{figure}[hbtp]
\begin{center}
\includegraphics[width=0.49\textwidth]{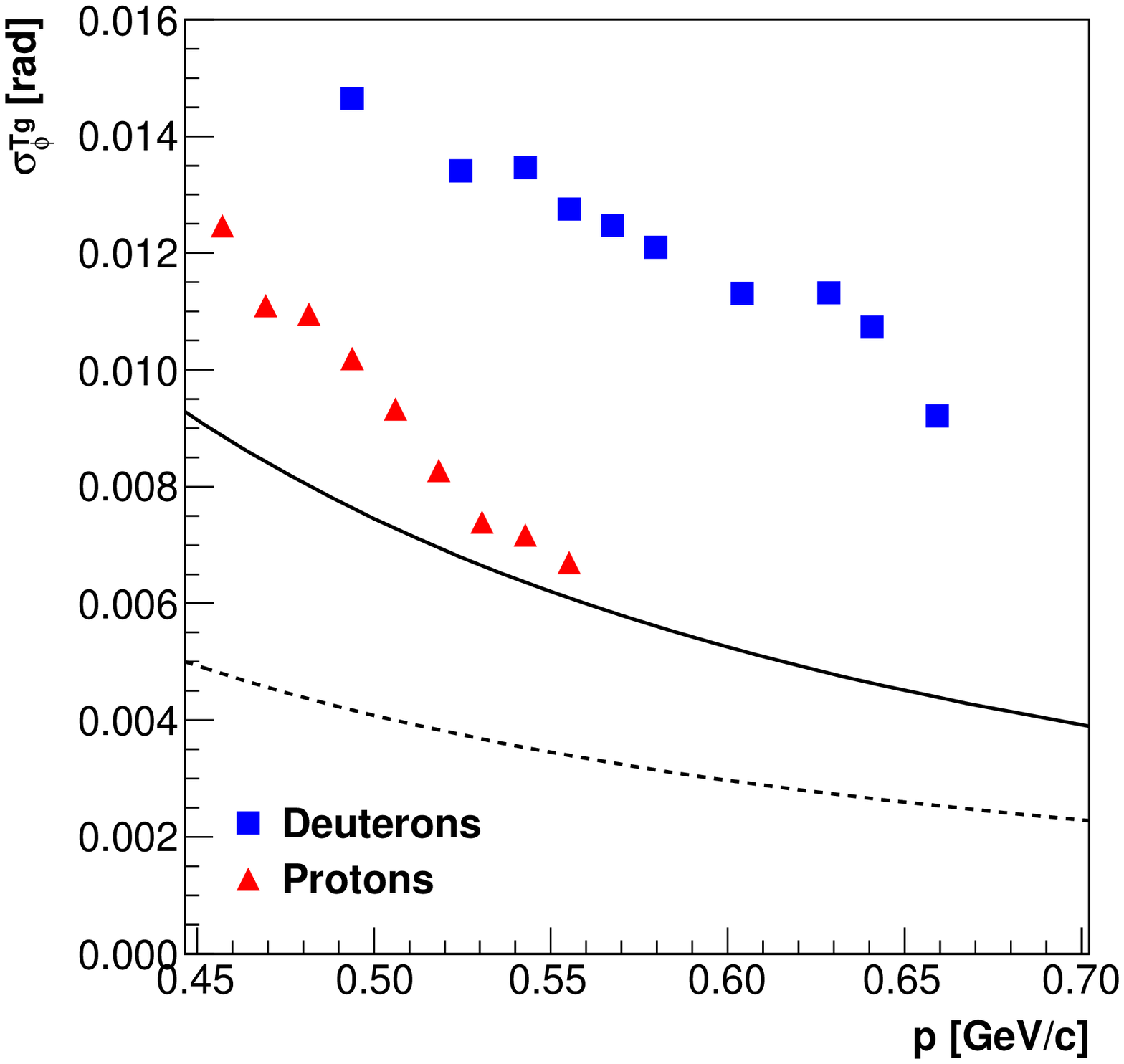}
\includegraphics[width=0.49\textwidth]{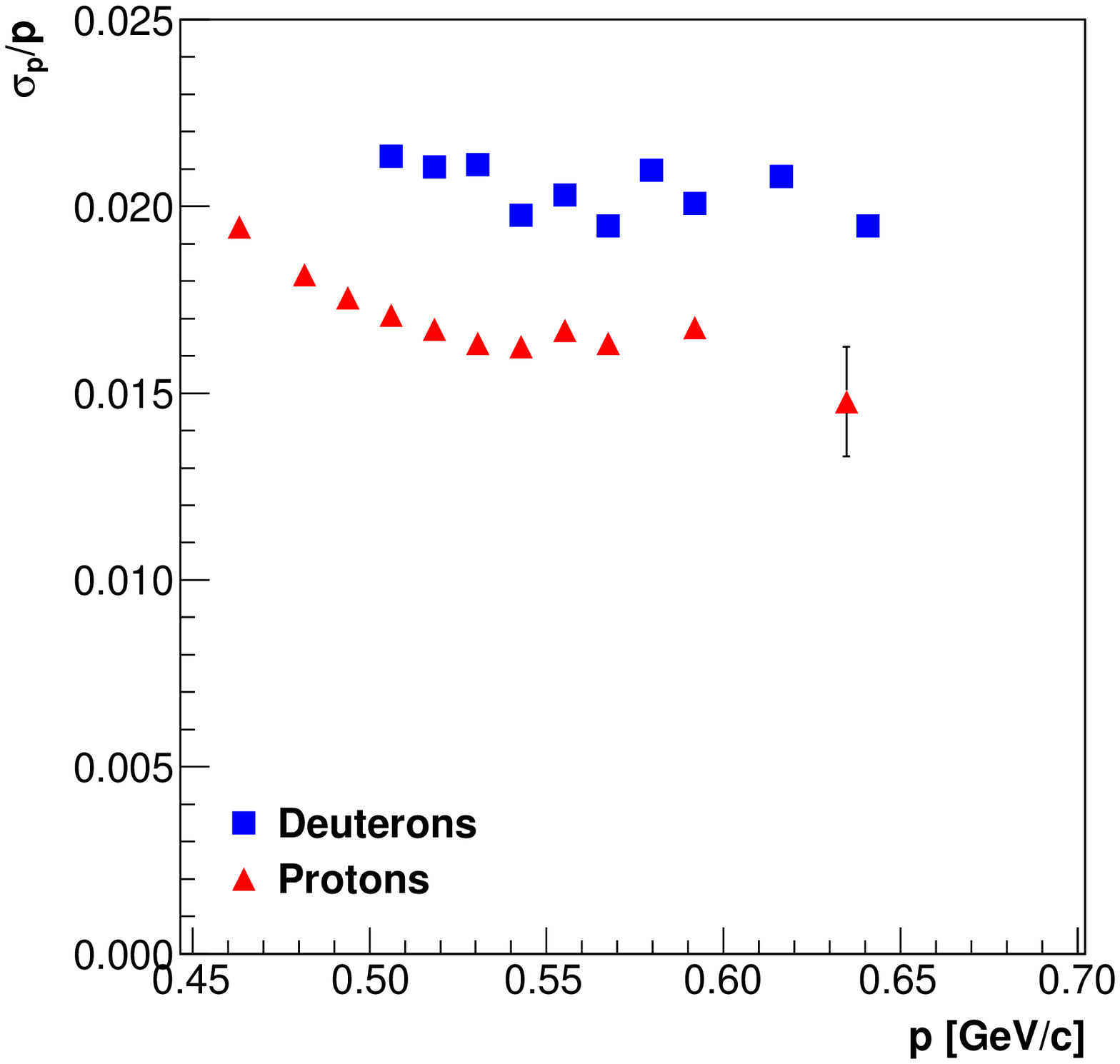}   
\vspace*{-3mm}
\caption{The absolute resolution of $\phi_{\mathrm{Tg}}$
and the relative momentum resolution as functions
of the momentum measured by BigBite, obtained by the SVD method. 
Irreducible multiple-scattering contributions, mostly
due to the air between the scattering chamber and MWDCs, 
are shown by full and dashed lines
for deuterons and protons, respectively. \label{ResolutionPlots}}
\end{center}
\end{figure}

The resolutions of $\theta_{\mathrm{Tg}}$ and $\phi_{\mathrm{Tg}}$ were
estimated by comparing them to the corresponding angles
as determined from the momentum transfer $\vec{q}$
in elastic scattering on hydrogen and deuterium.
The direction of $\vec{q}$ is given by the electron
kinematics and determined by the HRS spectrometer.
The corresponding HRS resolutions have been studied in \cite{GeJinPhD}.
Based on these values, the resolution of the reconstructed
$\vec{q}$ was estimated to be $6\,\mathrm{mrad}$
and $0.3\,\mathrm{mrad}$ for the vertical and horizontal angles,
respectively.  These contributions were subtracted in quadrature
from the calculated peak widths, yielding the final resolutions
attributable to BigBite.  The result for $\phi_{\mathrm{Tg}}$ is 
shown in Fig.~\ref{ResolutionPlots} (left).  The strong momentum 
dependence of the resolution is again caused by multiple scattering
in the target and the spectrometer.  Different resolutions
for deuterons and protons occur because the peak broadening
in multiple scattering strongly depends on the particle mass
(at a given momentum).  As before, the biggest contributions come
from the air.  In a typical kinematics
of the E05-102 experiment, the resolutions of 
$\phi_{\mathrm{Tg}}$ and $\theta_{\mathrm{Tg}}$ are 
$\sigma_{\phi_\mathrm{Tg}} \approx 7\,\mathrm{mrad}$ and
$\sigma_{\theta_\mathrm{Tg}} \approx 13\,\mathrm{mrad}$
for $0.55\,\mathrm{GeV}/c$ protons, and approximately 
$\sigma_{\phi_\mathrm{Tg}} \approx 11\,\mathrm{mrad}$ and
$\sigma_{\theta_\mathrm{Tg}} \approx 13\,\mathrm{mrad}$
for $0.6\,\mathrm{GeV}/c$ deuterons.
(Due to multiple scattering, these resolutions are clearly
much larger than the intrinsic MWDC resolutions mentioned
in Section~\ref{sec:BB}.)

The resolution of $\delta_{\mathrm{Tg}} = (p-p_\mathrm{c})/p_\mathrm{c}$
was also determined from elastic data by comparing the magnitude of $\vec{q}$
to the momentum reconstructed by BigBite.  The analysis
was done separately for the hydrogen and deuterium data sets.
Figure~\ref{ResolutionPlots} (right) shows
the relative momentum resolution $\sigma_{p}/p$
as a function of momentum.  The  relative momentum 
resolution  is approximately 
$1.6\,\%$ for   $0.55\,\mathrm{GeV}/c$  protons, and 
$2\,\%$ for $0.6\,\mathrm{GeV}/c$   deuterons.
Figure~\ref{MissingMassPlot} (right)  shows the absolute resolution
of $\delta_{\mathrm{Tg}}$.

\section{Summary}

We have described the optics calibration of the BigBite
spectrometer that was used to detect hadrons in the E05-102
experiment at Jefferson Lab.  While the methods have been
developed and applied to one spectrometer under very specific
physical conditions, the same procedures can be applied
to any spectrometer with a similar magnetic configuration
and acceptance.  

Two different approaches were considered: an analytical model
that treats BigBite as an ideal dipole and a matrix formalism.
The former approach results only in modest resolutions;
still, resolutions of a few percent can be achieved by
a suitable choice of parameters. The latter approach allows
for a more precise calibration.  Two numerical methods were used 
to determine the matrix elements, but the one based on singular
value decomposition delivered better and more reliable results.

The vertex resolution for protons was found to be  $1.2\,\mathrm{cm}$
at $0.55\,\mathrm{GeV}/c$ along the whole $40\,\mathrm{cm}$ target length.
The resolution deteriorates significantly at lower momenta
due to multiple scattering in the target, air, and detector material.
The corresponding angular resolution is $7\,\mathrm{mrad}$ for the
in-plane angle $\phi_{\mathrm{Tg}}$
and $13\,\mathrm{mrad}$ for the out-of-plane angle $\theta_{\mathrm{Tg}}$.
The angular resolution worsens at lower momenta due to
multiple scattering, with the effect more pronounced for deuterons.
The relative momentum resolution for  $0.55\,\mathrm{GeV}/c$ protons
(best case) has been estimated to be $1.6\,\%$. 

For $0.6\,\mathrm{GeV}/c$ deuterons (best case), we obtained
the resolutions of $2\,\%$ (momentum), $11\,\mathrm{mrad}$
($\phi_{\mathrm{Tg}}$), and $13\,\mathrm{mrad}$ ($\theta_{\mathrm{Tg}}$).

\section{Acknowledgments}

The Authors wish to thank Bogdan Wojtsekhowski for fruitful
discussions and valuable advice on the manuscript.

This work was supported in part by the U.S. Department
of Energy and the U.~S. National Science Foundation.
It is supported by DOE contract DE-AC05-06OR23177,
under which Jefferson Science Associates, LLC, operates
the Thomas Jefferson National Accelerator Facility.
The UK collaborators acknowledge the funding by
the UK Engineering and Physical Sciences Council
and the UK Science and Technology Facilities Council.

\vspace*{1cm}



%% References


\bibliographystyle{elsarticle-num}

\end{document}